%% file: main.tex
\documentclass{aa}  

\def\4u{4U1820-303}

\def\cstat{{\sc C-stat}}

\def\be{\begin{equation}} 
\def\ee{\end{equation}}

\def\ergscm2{erg s$^{-1}$ cm$^{-2}$}
\def\xspec{\texttt{XSPEC}}

\def\jaxspec{\texttt{jaxspec}}
\def\sixsa{\texttt{SIXSA}}
\def\xifu{\textit{X-IFU}}
\def\xrism{\textit{XRISM}}

\def\sbi{\texttt{sbi}}
\def\bxa{\texttt{BXA}}
\def\likelihoodemulator{\texttt{Likelihood\_emulator}}
\def\parameterretriever{\texttt{Parameter\_retriever}}
\def\bvapec{\texttt{bvapec}}

\usepackage{graphicx}
\usepackage{txfonts}
\usepackage{hyperref}
\usepackage{comment}
\usepackage{subfigure}
\begin{document} 

   \title{Simulation-based inference with neural posterior estimation applied to X-ray spectral fitting - III}
   \subtitle{Deriving exact posteriors with dimension reduction and importance sampling}

    \titlerunning{Neural networks for X-ray spectral fitting}
   \author{Didier Barret
          \inst{1}
          \and
          Simon Dupourqué\inst{1}}

   \institute{Institut de Recherche en Astrophysique et Planétologie, 9 avenue du Colonel Roche, Toulouse, 31028, France\\
              \email{dbarret@irap.omp.eu, sdupourque@irap.omp.eu}
              }

   \date{Received XXXX; accepted YYYY}

 
  \abstract
   {Simulation-based inference (SBI) with neural posterior estimation (NPE) can be used for X-ray spectral fitting both in the Gaussian and Poisson regimes, enabling users to rapidly derive approximated posteriors of the model parameters.} %
   {We investigate the capabilities of auto-encoders to reduce the dimension of X-ray spectra, such as the ones that will be provided by the X-ray Integral Field Unit (\xifu): the high-resolution X-ray spectrometer to fly on-board the European Space Agency \textit{NewAthena} space X-ray observatory. In addition, taking advantage of the known likelihood, we investigate importance sampling to refine the approximate posteriors.} %
   {We build an auto-encoder that compresses X-ray spectra into a low-dimensional latent space while preserving key spectral features. The auto-encoder is trained by minimizing a custom loss equal to the Cash statistic (\cstat) between the simulated and reconstructed spectra. A Neural Density Estimator (NDE) is then trained on the latent representations of the spectra. We use multi-round training for both the auto-encoder and the NDE. At each round, new spectra are drawn from a truncated proposal focused on the observation. Finally, when the NDE training has converged, the resulting approximate posteriors, conditioned at the observation, are refined via likelihood-based importance sampling. To evaluate the information content of the latent space, we introduce a diagnostic neural network trained to reconstruct the original spectral model parameters from the latent space. Additionally, we develop a specialized neural network that learns the likelihood function directly, enabling faster importance sampling and enhancing computational efficiency.} %
   {Reducing the dimension of \xifu-like X-ray spectra enhances the performance and efficiency of the neural posterior estimation. When combined with multi-round inference, our auto-encoder consistently outperforms other dimensionality reduction techniques such as Principal Component Analysis (PCA) and hand-crafted spectral summaries in terms of accuracy, and robustness. With each inference round, performance improves as the proposal distributions contract toward the observation. Following an importance-sampling correction, the resulting posterior distributions are statistically indistinguishable from those produced by nested sampling algorithms. On a standard multi-core laptop, the full pipeline, including simulations, dimension reduction, inference and subsequent importance sampling, achieves speedups exceeding an order of magnitude. Crucially, validation is based on real observational data, not just simulator outputs. In addition to mock \xifu\ spectra, we demonstrate successful applications to high-resolution \xrism-Resolve and lower-resolution \textit{NICER} and \textit{XMM-Newton} EPIC-PN observations, confirming the method  applicability across different instruments and spectral resolutions.
}
   {Simulation-based inference with neural posterior estimation on compressed X-ray spectra, when paired with likelihood-based importance sampling, yields posterior distributions that are indistinguishable from classical Bayesian results, offering an exact yet efficient alternative for X-ray spectral fitting. The \sixsa\ (\textbf{S}imulation-based \textbf{I}nference for X\textbf{}-ray \textbf{S}pectral \textbf{A}nalysis) python package available on GitHub is being update to include the auto-encoder and the importance sampling.}
   

   \maketitle
%

\section{Introduction}
X-ray spectral fitting is a cornerstone of high-energy astrophysics, providing access to the physical properties of compact objects, hot plasmas, and accretion flows through parametric modeling of count spectra folded through complex instrumental responses. Traditionally, parameter inference relies either on gradient-free minimization of fit statistics (e.g. $\chi^2$ or \cstat \citep{Cash1979ApJ...228..939C}) or on Bayesian sampling techniques such as Markov chain Monte Carlo (MCMC) \citep{hastings_monte_1970,vanDyk2001ApJ...548..224V,Goodman2010CAMCS...5...65G} and Nested Sampling (NS) \citep{Skilling2004AIPC..735..395S,10.1214/06-BA127,Buchner2014A&A...564A.125B,Buchner2021Ultranest,buchner_2023StSur..17..169B}. These techniques, albeit powerful, can be computationally expensive, sensitive to multi-modal or highly degenerate posteriors. They become increasingly difficult to scale as models and data grow in complexity, see \cite{Buchner2023arXiv230905705B} for a recent comprehensive review of statistical approaches in X-ray spectral analysis. Neural networks have emerged as an alternative, first applied by \cite{Ichinohe2018MNRAS.475.4739I} to high-resolution galaxy cluster spectra, and later by \cite{Parker2022MNRAS.514.4061P} to lower-resolution AGN data from \textit{NewAthena}. While offering comparable accuracy and a speed improvement of approximately three orders of magnitude after training, the latter approach lacked error estimates on spectral parameters. This limitation can be addressed within a Bayesian framework using Simulation-Based Inference with Neural Posterior Estimation (SBI-NPE; \cite{Papamakarios2016arXiv160506376P,Lueckmann2017arXiv171101861L}), which infers the posterior by training a neural network on synthetic spectra sampled from the prior. Unlike traditional Bayesian methods, SBI does not require an explicit likelihood and is thus applicable to complex models where a simulator is available \citep{Cranmer2020PNAS..11730055C, Deistler2025}. As with any likelihood-free approach, another strength of NPE is its ability to learn the likelihood function from compressed representations of the data. Dimensionality reduction serves multiple purposes: it isolates informative components, filters out noise, eliminates redundant features. This improves the training efficiency and prevents overfitting \citep[and references therein]{Deistler2025}.

We successfully applied simulation-based inference with neural posterior estimation for X-ray spectral fitting \citep[][hereafter BD24 and DB25]{Barret_2024AA,DupourqueBarret2025AA}. In BD24, we considered low-resolution spectra from the NICER instrument \citep{Gendreau2012SPIE.8443E..13G}. We showed that SBI-NPE performed on par with standard spectral fitting in both Gaussian and Poisson regimes, for both synthetic and real data, providing consistent parameter estimates and posteriors. In DB25, we investigated its applicability to high-resolution spectra expected from the upcoming \textit{NewAthena} X-IFU instrument \citep{Barret2023ExA....55..373B,peille-xifu-20205}, leveraging spectral compression and SBI likelihood-free nature.
Two spectral compression strategies were tested: (1) hand-crafted summary statistics (e.g. bin counts and ratios), and (2) neural network-based automated feature extraction. We showed that hand-crafted summary statistics yielded superior efficiency, producing posteriors comparable to those from exact inference. 

Following this up, we now investigate auto-encoders for compressing X-ray spectra and importance sampling for refining the posterior estimates. A key constraint throughout our study remains the ability to perform the entire analysis on a standard laptop (MacBook Pro, Apple M3 Max, 16 cores) within a practical duration: on the order of tens of minutes. This includes the time required to generate simulations, to train the auto-encoder, to obtain the truncated prior proposals, and to train the neural density estimator, considering that the posterior sampling time is negligible. To this, we now add the time to perform importance sampling, as introduced later. To minimize the end-to-end inference time, all components of the pipeline had to be optimized for efficiency.

The paper is organized as follows. In section \S \ref{sixsa_setup}, we describe the various components of the \sixsa\ pipeline. In \S\ref{results} we present the results of three test cases, involving spectral models of increasing complexity, before demonstrating the applicability of \sixsa\ to real \xrism-Resolve data. This precedes a discussion in \S\ref{discussion} and the conclusions 
\S\ref{conclusions}. In Appendices A-C, we report further results on \sixsa\ applied to mock X-IFU data and real \textit{NICER} and \textit{XMM-Newton} data.
\section{The \sixsa\ pipeline}
\begin{figure*}
        \centering
    \includegraphics[width=\linewidth]{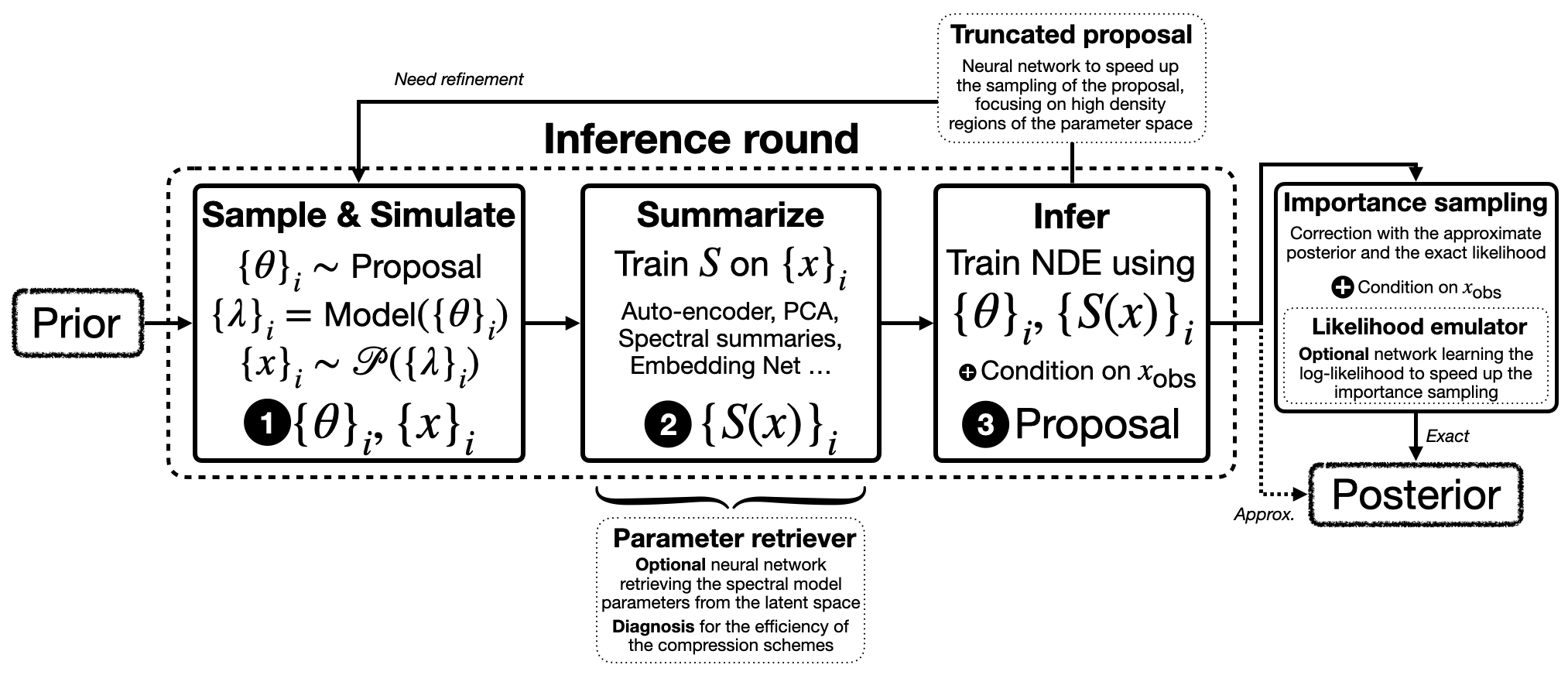}
    \caption{The \sixsa\ pipeline: The process begins with sampling parameters $\{\theta\}_i$ from a proposal distribution, followed by generating synthetic observations $\{x\}_i$ (including Poisson statistics), by passing these parameters to the spectral model. These spectra have their dimension reduced using various summarization techniques such as Principal Component Analysis (PCA), spectral summaries, or neural architectures like embedding networks and auto-encoders, yielding $\{S(x)\}_i$. $\{S(x)\}_i$, along with the corresponding parameters $\{\theta\}_i$ are used to train a neural density estimator (NDE). The  inference round is delimited with a dashed line. An optional parameter retriever neural network may be trained to learn the mapping between the latent space and the model parameters, aiding interpretation. For the observation, denoted as $x_{\rm obs}$, a truncated proposal network selectively focuses sampling on high-density regions of the parameter space. At each round, an approximated posterior can then be generated. Likelihood-based importance sampling can be applied to the approximated posterior. A likelihood emulator can also be used to approximate with high accuracy the true likelihood and accelerate importance sampling. This iterative process leads to the final posterior distribution for $x_{\rm obs}$.}
    \label{fig:sixsa_pipeline}
\end{figure*}

The different components of the \sixsa\ pipeline are represented in Figure \ref{fig:sixsa_pipeline}. Hereafter, we describe each of them, starting with the way the spectral simulations are performed, and following the order in the figure\footnote{The building blocks of the \sixsa\ (Simulation-based Inference for X-ray Spectral Analysis) pipeline are being released as a python package available on GitHub : \url{https://github.com/renecotyfanboy/bsixsa}.}. 

\label{sixsa_setup}
\subsection{Sample and simulate}
 Our ultimate objective is to develop an alternative technique for exploiting high-resolution X-ray spectra that is robust, reliable, accurate, and computationally efficient. As in Paper II, we first focus on high-resolution X-ray spectra expected from the \xifu\ instrument onboard \textit{NewAthena} \citep{Barret2023ExA....55..373B,peille-xifu-20205}. For our analysis, we adopt the latest available response files,\footnote{Response files available at: \url{https://x-ifu.irap.omp.eu/en/resources/for-the-community}} which serve as the current baseline for \textit{NewAthena} in preparation for the adoption of the mission in the science program of the European Space Agency. These are the reference files intended for use in the so-called Red Book. We consider the open configuration of the filter wheel (i.e., no filter), which maximizes the instrument effective area below approximately 1 keV.
 
The spectral models considered below involve multiplicative and additive components. As in BD24 and DB25, the normalization parameters of additive components, which typically span several orders of magnitude and represent multiplicative scaling of model flux, were assigned log-uniform (Jeffreys) priors. For shape parameters such as photon index, temperature, or absorption density, which vary over narrower and more physically constrained ranges, we adopted uniform priors to reflect equal prior belief across the specified linear interval (see the discussion about choosing priors in \cite{Buchner2023arXiv230905705B} and in \citep{Buchner2014A&A...564A.125B}). 
For each spectral model, we use the Python interface to the \xspec\ spectral-fitting package (\texttt{PyXspec}, \cite{Arnaud1996ASPC..101...17A}) to fold it through the \xifu\ response. We simulate spectra with total counts ranging from approximately 50,000 to 150,000 over the 0.2–12 keV energy range. For each test case presented below, we simulate one mock spectrum defining our target observation denoted $x_{\rm obs}$ in Figure \ref{fig:sixsa_pipeline}. We adopt optimal binning for $x_{\rm obs}$, following the approach of \citet{Kaastra2016A&A...587A.151K}. By accounting for count statistics and energy distribution, this typically reduces the number of spectral bins to between 2,000 and 3,000, yielding a compression factor of $\sim10$. For a given $x_{\rm obs}$, all spectra simulated (step 1 of Figure \ref{fig:sixsa_pipeline}) share the same binning scheme. Background contributions are not included in this analysis (see DB25 for a detailed treatment of background handling).  Our simulations are performed using \xspec\ version 12.15. Notably, starting from version 12.14.1, a new mechanism for setting non-default model parameters has been introduced, significantly improving computational efficiency. This enhancement accelerates the generation of \xifu\ mock spectra by approximately a factor of two, with 10,000 spectra now typically produced in about 15 seconds for most models (for spectra of $\sim 2,000-3,000$ bins). To achieve this performance, we first load the target observation along with its data grouping, then set all model parameters in parallel. Through internal \xspec\ computations, we retrieve in a single operation the \cstat\ value (hence, the log-likelihood), the input model, and the folded model. We then apply Poisson noise externally to the folded model. Importantly, this approach avoids the use of the \texttt{PyXspec} \texttt{fakeit} command and further eliminates any need for disk I/O operations. In our setup, the spectral model evaluation constitutes the main computational bottleneck. 

In our previous work, we emphasized the benefits of constraining the prior range before entering the inference loop, as a mean to accelerate convergence. In the present study, however, our focus is on enhancing the robustness of the technique. To this end, we extend the prior to ensure that $x_{\rm obs}$ lies near the center of the prior predictive distribution. This is achieved by adjusting the bounds of the normalization parameters for the additive components. This involves either lowering the minimum or raising the maximum bound by a given factor. This operation is performed iteratively, and at each iteration, new spectral model parameters are drawn from the updated (and broader) priors. The end product is a training set in which the histogram of the total number of counts per simulated spectrum is distributed symmetrically around the total number of counts of $x_{\rm obs}$, within say $\sim $10\% (see the left panel of Figure \ref{fig:test-case-I-prior-coverage} for an illustration).
 
 \subsection{Summarize} As discussed in DB25, for simulation-based inference to work with \xifu-like spectra with thousands of spectral bins, it is essential to reduce dimensionality through summary statistics. We compute one set of summary statistics per spectrum. As emphasized by \cite{Deistler2025}, summary statistics \(S\) should retain as much information as possible about the parameters \(\theta\), ensuring that the posterior derived from the summary statistics, \(p(\theta \mid S(x))\), closely approximates the  posterior \(p(\theta \mid x)\).  

Here, we provide some details about the dimension reduction techniques considered in this paper. 
\subsubsection{An auto-encoder tailored for X-ray spectra}
Auto-encoders are a class of neural networks used for unsupervised learning, particularly effective for dimensionality reduction and feature extraction \citep[e.g.][]{Hinton2006Sci...313..504H}. They comprise an encoder, which compresses input data into a lower-dimensional latent space, and a decoder, which reconstructs the input from this compressed representation. This architecture enables the model to capture the most important features of the data, which is necessary for good reconstruction from the low-dimensional latent space. Training is typically guided by a custom loss function that minimizes the reconstruction error, allowing the auto-encoder to learn efficient and informative representations.

We have developed an auto-encoder tailored for X-ray spectral data. The same auto-encoder architecture will apply to all the spectral models considered below. The encoder is composed of a sequence of fully connected layers, each followed by batch normalization, a GELU activation function \citep{hendrycks2023gaussianerrorlinearunits}. This structure progressively reduces the input dimensionality, mapping the spectra into a latent space. The decoder mirrors this architecture in reverse. To account for the Poisson statistics of the simulated spectra, training is performed by minimizing the Cash statistic (\cstat) between each simulated spectrum and its reconstruction. This is equivalent to a Poisson reconstruction loss. To facilitate convergence and achieve more stable learning, the spectra are log-transformed using $\log(1 + c_i)$, where $c_i$ is the count in spectral bin $i$, followed by standardization (zero mean and unit variance); the inverse transformation being applied before the loss computation. The training process incorporates standard optimization techniques including gradient clipping, learning rate scheduling, and early stopping to enhance convergence and prevent overfitting. We use a batch size of 1024, a maximum of 500 training epochs, and early stopping patience of 20. The auto-encoder is trained at each inference round, see Figure \ref{fig:sixsa_pipeline}. At each new round, training of the auto-encoder resumes from where it left off in the previous round by warm-starting from the saved checkpoint. This process restores both the model parameters and the optimizer state, allowing training to continue seamlessly on spectra sampled from the updated proposal distribution, as illustrated in Figure~\ref{fig:sixsa_pipeline}. This means that training in the first round takes significantly longer than in subsequent rounds, as later rounds benefit from the accumulated learning.

For the models considered here, a latent space dimension of 64 provided the best balance, yielding efficient training of the auto-encoder and stable performance in the subsequent training of the neural density estimator. Increasing the latent dimension to 128 did not improve the auto-encoder training and could lead in some cases to clear signs of overfitting, as evidenced by the validation loss during the neural density estimator training. Although we consider latent space dimensions to powers of two, this choice is arbitrary and not motivated by theoretical considerations.

\subsubsection{Principal Component Analysis and spectral summaries}\label{subsubsec:pca} Principal Component Analysis (PCA) is a widely used dimensionality reduction technique, particularly in machine learning pipelines \citep{Jolliffe1986pca..book.....J}. It was applied by \cite{Parker2022MNRAS.514.4061P} to reduce the dimensionality of \textit{NewAthena} Wide-Field Imager X-ray spectra before training a neural network to
predict their spectral parameters. The PCA projects the data onto a set of orthogonal principal components, ordered by the fraction of variance they capture. The procedure involves standardizing the data, computing the covariance matrix, and performing eigen decomposition to extract these components. PCA, however, relies on linear assumptions and may fail to preserve the most informative features. In our case, we reduce the dimensionality of the standardized spectra while retaining 99.5\% of the variance. The achieved reduction depends on the training set, and the method becomes less effective as the parameter space narrows: the number of components required to capture the variance approaches the number of spectral bins, leading to potential overfitting during the NDE training.

A set of X-ray spectra can also be reduced to a limited number of summary statistics, such as the mean number of counts per bin, the standard deviation, the total counts, the skewness, the entropy, and specific percentiles (e.g. 90th and 95th) or the inter-quartile range (IQR), which capture key characteristics of the distribution. Additional features can be derived from adjacent energy intervals, including the sum of counts, excess counts, energy-weighted mean counts, count-weighted mean energies, hardness ratios ($C_{i+1}/C_i$), and differential ratios ($(C_{i+1}-C_i)/(C_i-C_{i-1})$), providing complementary information on spectral shape and variability across energy bands. These ratios contribute to the parameter recovery for more complex models, such as those with two \bvapec\ components (DB25). 

As discussed in App. B of DB25, the summary statistics of the observations should be covered by the summary statistics of the simulated spectra to ensure that the neural density estimator can interpolate between the summaries on either side of the observation. Statistics that do not cover the observable can be removed from the training sample to improve the speed and stability of the gradient descent. This approach, however, may change the number of summaries retained from round to round, requiring the density estimator to be retrained from scratch, which is our default procedure. Although these summaries may not capture fine spectral details, keeping their number small ($\leq 100$) minimizes the risk of overfitting during neural density estimator training. This makes the method particularly effective for continuous spectra without distinct features.

For the spectral summaries considered here, we use 10 adjacent energy intervals in addition to the broad-band summaries, in order to limit overfitting.
\subsubsection{A Parameter retriever: retrieving the spectral model parameters from the latent representation of the spectra}
Before feeding the compressed representation of the spectra into the neural density estimator for training, it is useful to exploit the dimensionality reduction as an intermediate step. Specifically, we test whether a neural network can learn the mapping between the model parameters and the latent representation of the spectra, a task infeasible with the raw spectra containing several thousand bins. This provides an initial indication of whether the compression scheme successfully retains the main features of the training sample (hence the model parameters) and whether all model parameters are equally constrained by the observation. 

For this purpose, we design a multi-layer perceptron (MLP) composed of fully connected layers interleaved with ReLU activation functions. The loss function is the Mean Squared Error (MSE), which is well suited for regression tasks. Although the spectral parameters span different physical scales, we standardize them prior to training, ensuring that each contributes equally to the loss in the standardized space. The input summary statistics are also standardized to stabilize and accelerate the training process. The MLP is trained using the Adam optimizer. In addition, a learning rate scheduler is employed to reduce the learning rate gradually during training. Hereafter, to avoid confusion, we refer to this neural network as the \parameterretriever. By default, it consists of three hidden layers with 128, 64, and 32 units, respectively. The model is trained over 500 epochs, with a batch size of 128 and an initial learning rate of $10^{-3}$. Early stopping is applied with a patience of 20 epochs to prevent overfitting. The \parameterretriever\ is not required in the \sixsa\ pipeline. It can be trained and used within each inference round. As expected, the performances of the \parameterretriever\ improve and stabilize after the first inference round is completed, when the proposal from which thetas are generated has shrunk compared to the initial (broader) prior.

\subsection{Infer}
Once the simulated spectra are summarized, we can now perform the inference. We use the \texttt{Simulation-Based Inference (sbi)} Python package~\citep{Tejero2020JOSS....5.2505T}, and among its available methods, we adopt Neural Posterior Estimation (NPE)~\citep{Papamakarios2016arXiv160506376P,Lueckmann2017arXiv171101861L,greenberg2019automaticposteriortransformationlikelihoodfree}, which is the most widely used technique within the \texttt{sbi} framework. NPE trains an inference network to directly approximate the posterior distribution \( p(\theta \mid x) \), using samples \((\theta, x)\) drawn from the joint distribution \( p(\theta, x) \). Unlike methods that focus on predicting a single point estimate for the parameters, NPE trains the network to predict the parameters of a probability distribution over \(\theta\), conditioned on the observed data \(x_o\).  Such networks are trained by minimizing the conditional negative log-likelihood of simulated parameters under the neural posterior estimator, corrected with a proposal function depending on the round as suggested by \cite{greenberg2019automaticposteriortransformationlikelihoodfree}, see also \cite{deistler2022truncated} for broad introduction on these topics. We employ multiple-round inference (MRI), using  typically five rounds. The size of the training sample varies from one test case to the other: it varies from 10,000 in the test case I, up to 50,000 in the test case III, depending primarily on the model complexity. There is no requirement to keep the training sample size constant from one inference round to the other, but we have considered it fixed hereafter. MRI is tailored for the analysis of a single observation, but the same architecture could be extended to amortized inference (applicable to multiple observations) at the cost of significantly increasing the simulation budget, depending on the diversity of the spectral properties across the dataset.

We adopt a Masked Autoregressive Flow (MAF) as the density estimator, using 10 transformations with 100 hidden units each \citep{papamakariosMaskedAutoregressiveFlow2017}. We also evaluated alternative architectures, including Neural Spline Flows (NSF), Mixture Density Networks (MDN), Masked Auto-encoders for Distribution Estimation (MADE), MAFs with Rational Quadratic Spline (RQS) transformations, and a separate MAF implementation via ZUKO (instead of NFLOWS). A more in-depth investigation of the performance of the various density estimators is beyond the scope of this paper; however, the selection of MAF was motivated by comparisons of the validation loss, training stability, and the shape of the posteriors, after a given number of inference rounds. The conditional density estimator for the posterior is trained at each inference round, see Figure \ref{fig:sixsa_pipeline}. At each new inference round, it is retrained from scratch, which is necessary when the dimensionality of the compressed data changes between rounds, as with the PCA. When the dimensionality remains fixed, not retraining from scratch yields consistent results, though the training time tends to increase slightly with each successive round. We consider a learning rate for the Adam optimizer of $5\times10^{-4}$, a batch size of 256, a validation fraction of 0.2, and patience of 20. Independent z-scoring (standardization) is applied to both $\theta$ and $x$, as defined earlier. Following \citet{deistler2022pnas2207632119}, we apply prior truncation at each inference round by rejecting samples that fall outside the $10^{-3}$ quantiles of the support of the approximate posterior. For a comprehensive practical guide on simulation-based inference, we refer to \citet{Deistler2025}, whose recommended best practices were adopted in this work. In particular, we double the training sample size in the first round to improve initial posterior estimation.
 
\subsection{Importance sampling} \label{sir} 
Unlike traditional Bayesian inference, simulation-based inference requires only access to simulations from the model and does not rely on evaluating the likelihood function explicitly. However, in our case, the likelihood is known. It can then be used to refine the posterior estimates. Importance sampling is now applied in various fields, e.g. for inferring atmospheric properties of exoplanets \citep{Gebhard2023arXiv231208295G, Gebhard_2025A&A...693A..42G}, for accurate and reliable gravitational wave inference \citep{Dax2023PhRvL.130q1403D} (see \cite{Tokdar_10.1002/wics.56} for a review on importance sampling). Here, for the first time, it is introduced in the context of \sixsa.
\subsubsection{Methodology}
In our set-up, the NPE yields an approximate posterior
\(q_\phi(\theta \mid x)\).
When both the prior \(p(\theta)\) and the likelihood \(p(x \mid \theta)\) can be computed, this neural posterior can be refined by importance sampling.
Given the observed data \(x_o\), draw samples \(\{\theta_i\}_{i=1}^N \sim q_\phi(\theta \mid x_o)\) and attach weights
\[
w_i \;\propto\; \frac{p(\theta_i)\,p(x_o \mid \theta_i)}{q_\phi(\theta_i \mid x_o)}\,.
\]

Intuitively, samples that are underestimated by the neural posterior relative to the true posterior receive larger weights and vice versa.

 \sbi\ implements two importance sampling methods, that we have implemented for testing purposes, given that this is the first time they are applied within \sixsa. 
 
 The first method is named \textit{Weighted Importance Sampling} (WIS): this returns samples from the proposal and the logarithm of the importance weights. The latter are converted into regular weights by first removing infinite values, subtracting the maximum, and then taking the exponential (this gives a pair $\theta_i, w_i$). Normalizing the weights \(\tilde w_i = w_i / \sum_j w_j\), one can define an effective sample size (ESS) also called the number of effective samples as : 
  \[
  \mathrm{ESS} \;=\; \frac{\big(\sum_i \tilde w_i\big)^2}{\sum_i \tilde w_i^2}\,,
  \]
  where small values indicate a poor match between ${q_\phi(\theta_i \mid x_o)}$ and the target (the sample efficiency is $\epsilon=$ ESS/$n$, where $n$ is the number of drawn samples and is $\in [0,1] $). As stated in \cite{Gebhard_2025A&A...693A..42G}, $\epsilon \gtrsim 1$ is considered a good value. If desired, resampling according to \(\tilde w_i\) enables to obtain the desired number of posterior samples. 
  
The second method is named \textit{Sampling-Importance-Resampling} (SIR). It directly returns samples by performing weighted importance sampling in batches. For each batch, it computes the normalized weights $\tilde{w}_i$ as described above, and then resamples indices with probabilities $\tilde{w}_i$ to obtain an unweighted sample. Given a batch size $b$ (i.e., the oversampling factor), one sample is drawn per batch according to its importance weight. To obtain $N$ posterior samples, a total of $N \times b$ likelihood evaluations are required. This approach may introduce a small bias due to the finite-sample nature of the resampling step. In practice, we observed a non-negligible bias in some runs, which was not corrected for certain parameters.

Importance sampling performs best when the true posterior is supported by the proposal distribution. As we will see below, even a moderately accurate neural posterior can be noticeably improved by reweighting, correcting residual mismatch and sharpening credible regions with minimal additional computation. Unless otherwise mentioned, within the \texttt{sbi} package, we have considered the weighted importance sampling method, considering between 200,000 and 400,000 likelihood estimates, to match typical sampling efficiencies. This becomes a significant contributor to the simulation budget, hence run time, but this is undoubtedly an acceptable price to pay to get accurate posteriors. 
\subsubsection{A likelihood emulator to speed up importance sampling}
For importance sampling, a large number of likelihood evaluations is required. Following \cite{Graff_2012MNRAS.421..169G}, we are considering a neural network to learn the likelihood function. Specifically, within the range of the approximate posteriors, a neural network can efficiently learn the mapping between the spectral model parameters and the likelihood using a limited training sample (tens of thousands). To enhance training stability and convergence, both the input features (model parameters) and the target variable (\cstat) must be standardized. We employed unit variance standardization and observed that within the tests performed in this paper, it resulted in higher prediction accuracy compared to alternative methods, such as \texttt{MinMax} or \texttt{robust} scaling for instance. To further improve the prediction accuracy, one can remove a small fraction (0.1-1\%) of the lowest log-likelihoods (and the corresponding parameter samples), as they cover a range of values not used during the importance sampling correction. The neural network architecture comprises three hidden layers with 128, 64, and 32 units, respectively. Each layer is followed by the Gaussian Error Linear Unit (GELU) activation function. The network concludes with a single-unit linear output layer suitable for regression tasks. Model training was performed using the Adam optimizer with a learning rate of $10^{-3}$, a batch size of 128, and a maximum of 1,000 epochs. To mitigate overfitting, early stopping with a patience of 50 epochs was implemented. Additionally, dynamic learning rate reduction was applied during training to further enhance performance and generalization. Here, we refer to this neural network as a \likelihoodemulator. Once trained (within a couple of minutes), the evaluation of the approximate likelihood function is done with within ten seconds for hundreds of thousands of spectral model parameters, enabling the importance sampling to be performed readily. However, caution is required for multi-modal posterior distributions because the parameter range over which to learn the likelihood may be too large. For this reason, we would recommend checking the accuracy of the predictions of the \likelihoodemulator\, and in doubts always prefer the exact, though more computationally expensive, computation. Yet for all the models, considered below, training a \likelihoodemulator\ with 50,000 samples led to sufficiently accurate likelihood estimates.

\subsection{Posterior comparison}
At the end of the \sixsa\ pipeline (see Figure \ref{fig:sixsa_pipeline}), posteriors are derived. For posterior comparisons, we perform \xspec\ MCMC sampling using the Goodman-Weare algorithm \citep{Goodman2010CAMCS...5...65G} with a burn-in phase of 25,000 steps, a chain length of 500,000, 64 walkers, and 16 cores. We apply Jeffreys/LogUniform priors on the model normalization parameters and uniform priors on all other parameters, with Bayesian inference enabled. For \bxa, we adopt the same priors as used for \xspec, employing 3,000 live points and the recommended reactive nested sampler \citep{Buchner2014A&A...564A.125B}. The number of live points was chosen to ensure that for the spectral models and priors considered in the paper, a single \bxa\ run would yield at least 50,000 posterior samples: a number that was requested also to \sixsa. Reducing the number of live points would decrease the \bxa\ runtime at the expense of reducing the number of posterior samples, see \cite{buchner_2023StSur..17..169B} for a discussion on the  relationship between number of live points, dimensionality and computational cost.

\section{Results}
After presenting the \sixsa\ pipeline, we now report the results obtained from test cases involving spectral models of progressively increasing complexity.
\label{results}
\subsection{Test case I : a narrow line above a smooth continuum}
\begin{figure}[!h]
    \centering
    \includegraphics[width=0.51\linewidth]{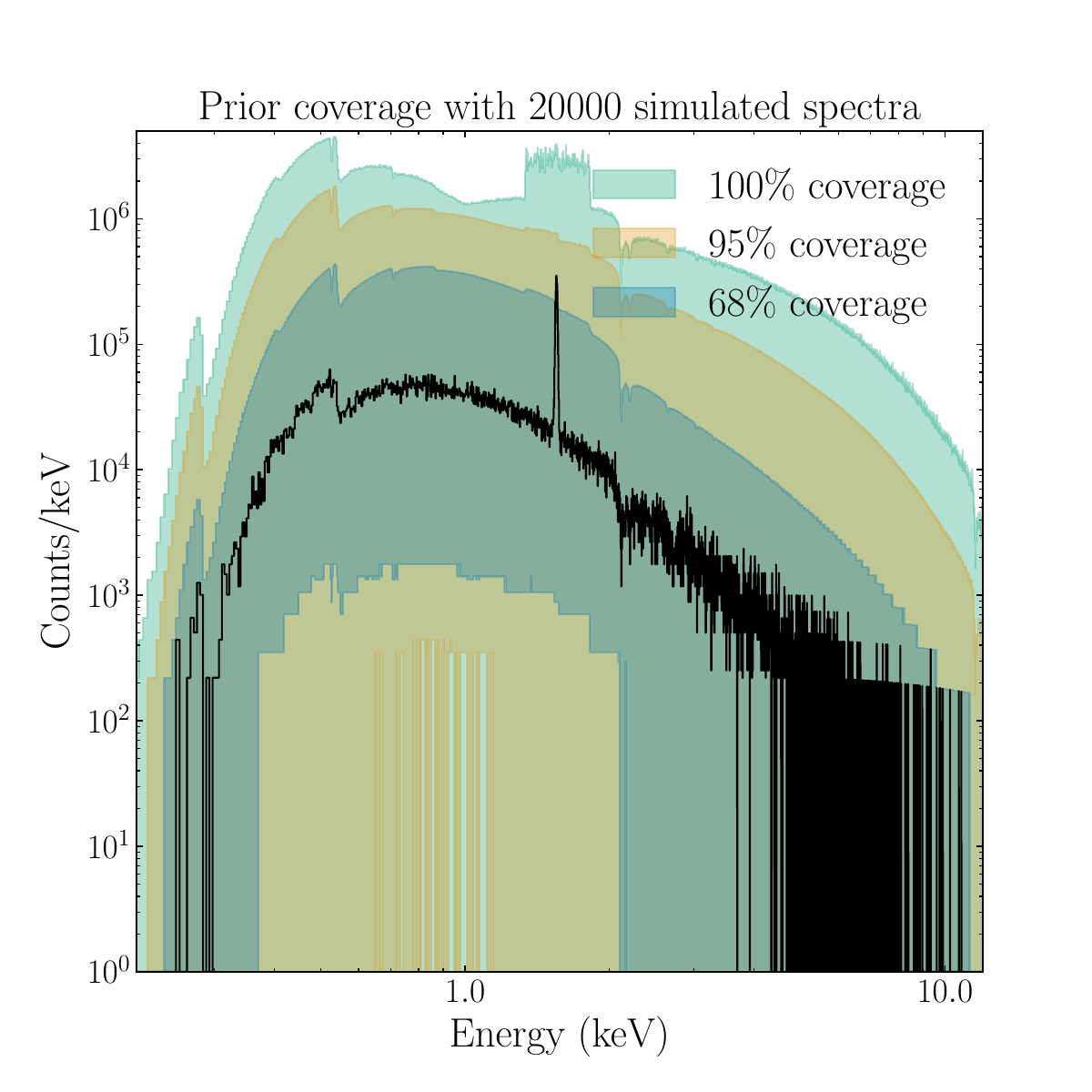}\includegraphics[width=0.51\linewidth]{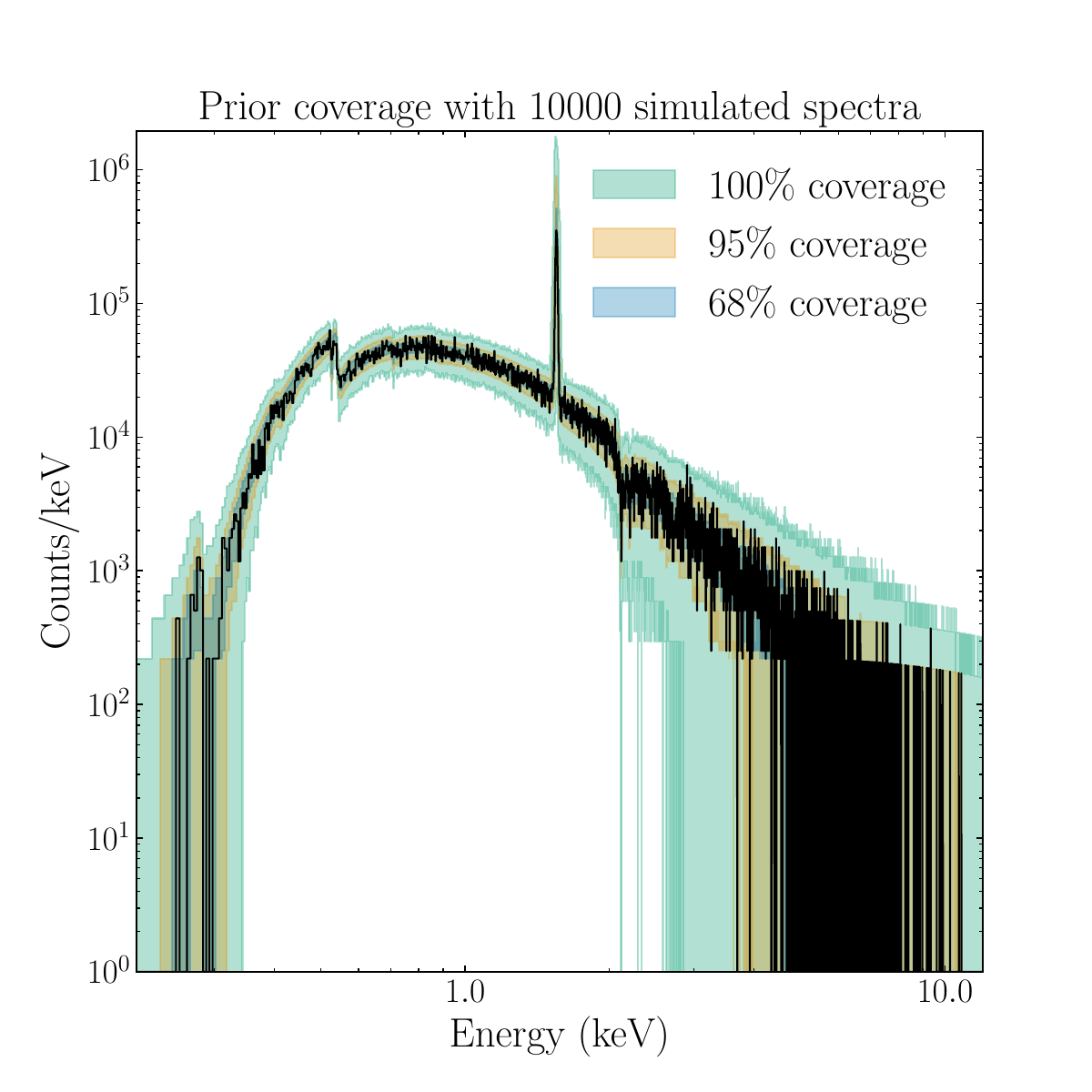}
    \caption{Left: The initial prior coverage of the targeted observation (black line), obtained with 20,000 spectra: twice the number used in the subsequent round. The range of the initial prior has been expanded to ensure a similar number of training sample spectra have total number of counts below and above the targeted observation (this ensures the observation to be centered). Right: The prior coverage in round 2, based on 10,000 simulations. The allowed parameter space has shrunk considerably. This prior proposal could be used as input for running \bxa, significantly speeding up the inference by reducing the parameter space that needs to be explored.
}
    \label{fig:test-case-I-prior-coverage}
\end{figure}

\begin{figure}[!h]
    \centering
    \includegraphics[width=0.4975\linewidth]{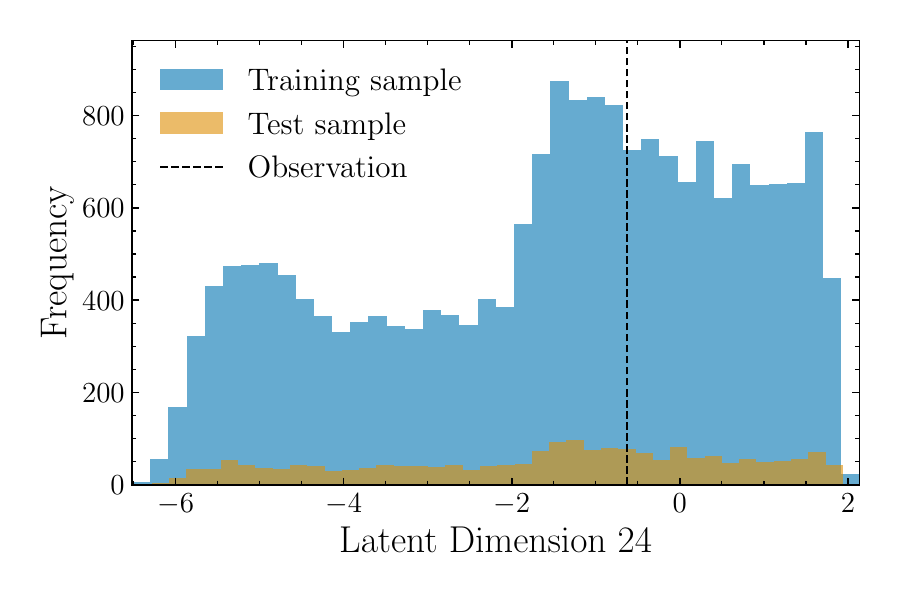}\includegraphics[width=0.4975\linewidth]{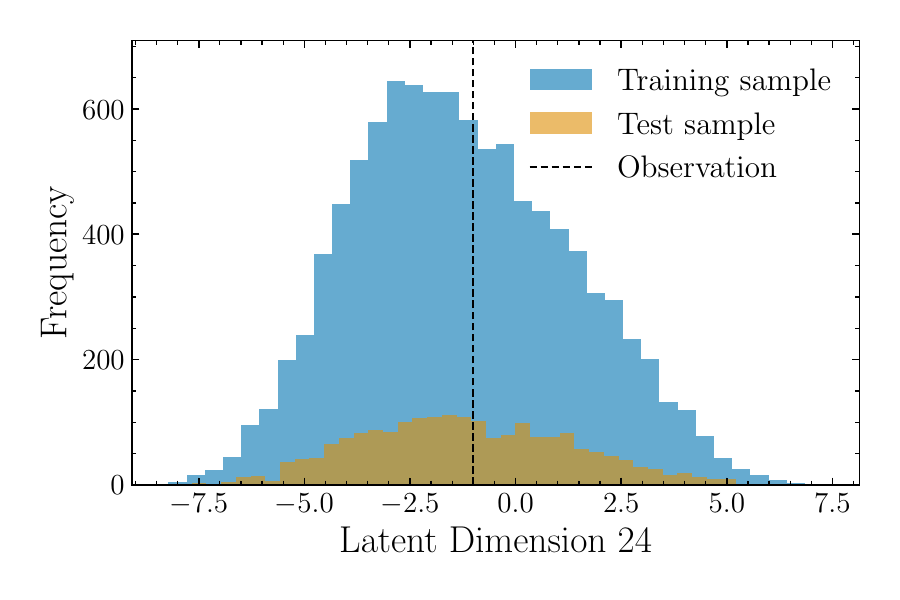}
    \caption{Left: The histogram of the 24th of the 64 latent space dimensions for the auto-encoder training and test samples as derived from the initial prior shown in the left of Figure \ref{fig:test-case-I-prior-coverage}. The 24th latent dimension of the observation is materialized by the dashed vertical line. Right: The same histogram but derived from spectra simulated from the prior proposal generated after the single round of MRI (shown in the right of Figure \ref{fig:test-case-I-prior-coverage}). The shape of the histogram evolves towards a gaussian-like shape and is better centered around the observation.}
    \label{fig:latent-space-dimension}
\end{figure}

 We are interested in probing dimension reduction techniques in the presence of lines in the X-ray spectrum, as those may be affected by the loss of information related to the dimension reduction. We start with a very simplistic case in which the spectrum consists of an absorbed power law and a narrow line of fixed energy but varying normalization, width and redshift, which we are interested in (see Figure \ref{fig:test-case-I-prior-coverage} for the initial prior coverage of the targeted observation and Table \ref{tab:parameter_limits_caseI} for the range covered by the spectral model parameters). 
\input{figures/caseI/parameter_limits}
We consider 5 rounds of inference and a training set of 10,000 simulations per round, except the first round for which we double the number of simulations. The model being analytical, evaluating the model to fold with the \xifu\ response is extremely fast, hence the simulations of ten thousand spectra is readily done within $\sim 15$ seconds (16 cores). Such a model, with a line above a smooth continuum, is expected to challenge the dimension reduction via the PCA and the spectral summaries. 
We use the auto-encoder with one hidden layer of width \(D/2\) (where \(D\) is the number of spectral bins) and a 64-dimensional output, yielding a compression ratio of approximately \(D/64\) (e.g., \(\sim 50\times\) for \(D\approx 3200\)). 
In Figure \ref{fig:latent-space-dimension}, we show the histogram of the $24^{\text{th}}$ of the 64 latent space dimensions as derived from the auto-encoder run on a sample of spectra generated from the initial prior and second round prior, as shown in Figure \ref{fig:test-case-I-prior-coverage}. As can be seen, already at the second inference round, the shape of the histogram evolves towards a gaussian-like shape and is better centered around the observation.

\begin{figure}[!h]
    \centering
    \includegraphics[width=0.75\linewidth]{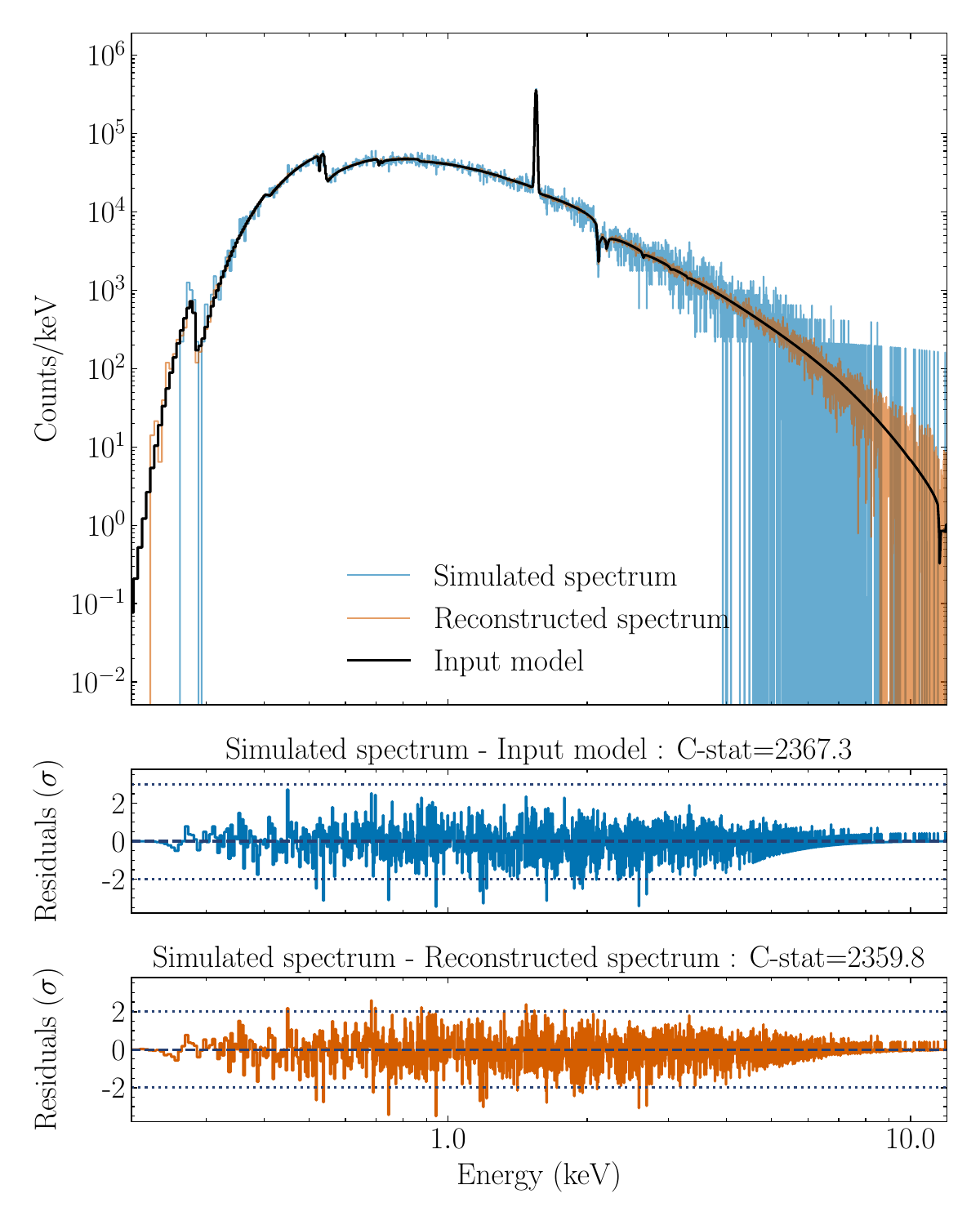}
    \caption{A random simulated spectrum including Poisson noise (blue), with its input model (black line). The spectrum reconstructed by the decoder part of the auto-encoder is shown in orange. The residuals, expressed in $\sigma$, between the simulated spectrum and both the input model and the reconstructed spectrum are shown in the bottom panels. The corresponding \cstat\ (without minimization) is listed for indication.}
    \label{fig:auto-encoder-reconstruction}
\end{figure}

As discussed above, the auto-encoder includes a decoder that reconstructs the spectrum from its latent representation (hereafter we name the latter spectrum, the reconstructed spectrum). The auto-encoder performance can be assessed by comparing the \cstat\ of simulated spectra versus their input models (without minimization) and their reconstruction  (computed from the decoder section of the auto-encoder); good reconstructions should yield comparable \cstat\ with the actual ground-truth spectrum.  Figure \ref{fig:auto-encoder-reconstruction} provides an illustrative example. As can be seen, the \cstat\ are comparable, indicating that the auto-encoder has captured the overall shape of the simulated spectrum, but also the narrow line, leaving no apparent residuals around it.

\begin{figure}[!h]
    \centering
    \includegraphics[width=0.7\linewidth]{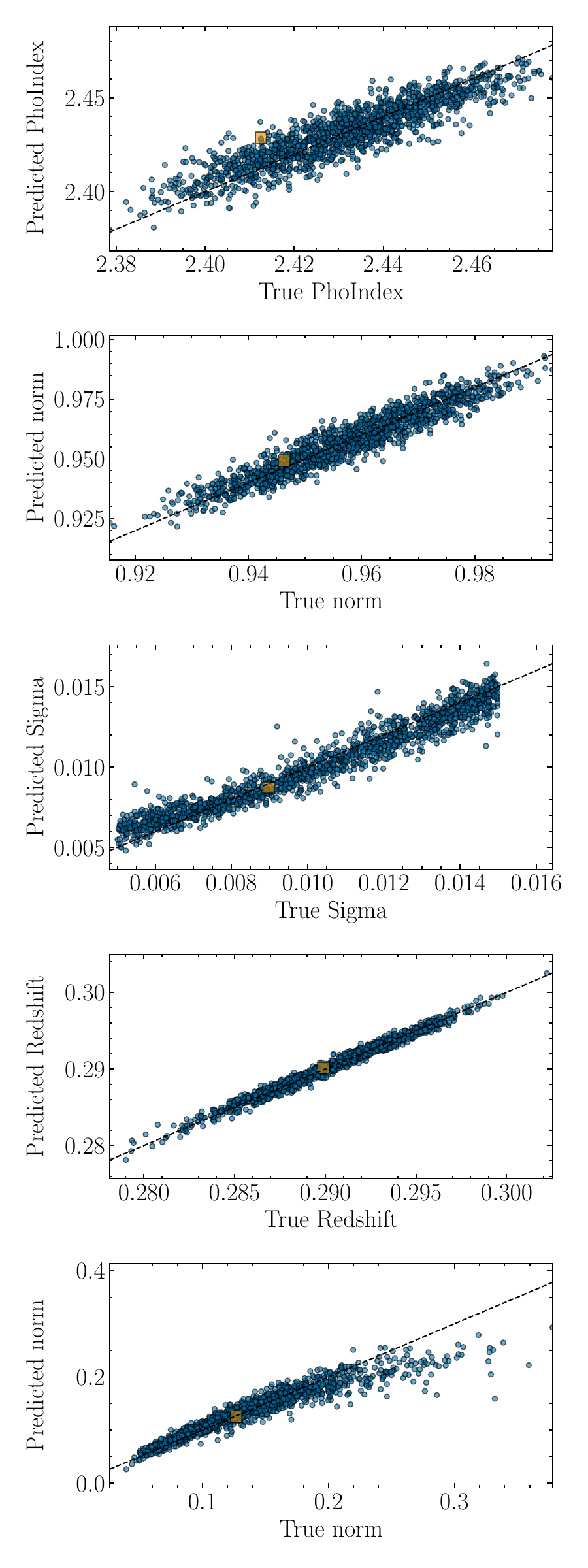}
    \caption{The mapping by the \parameterretriever\ between the model parameters and the latent space representations (64 dimensions) of 2,000 test spectra at the second round of inference. In each panel, the true parameter values are shown on the x-axis, and the predicted values on the y-axis. This result indicates that the autoencoder has successfully captured the relevant information from the simulated spectra. Moreover, the strong alignment between predicted and true values confirms that the observation is sensitive to all five model parameters. The target observation is marked with a square symbol for reference.}
    \label{fig:parametermapper}
\end{figure}

Figure \ref{fig:parametermapper} illustrates the predictions of the \parameterretriever\ against the five input model parameters. A close match between predicted and true values is observed, indicating that the observation is sensitive to all five parameters. In contrast, if the observation were insensitive to a given parameter, the reconstruction would appear as a flat line centered around the mean of the parameter prior range (see Appendix \ref{Appendix A}). In such a case, the corresponding parameter could be safely frozen during the fit. Although not strictly necessary, this type of sanity check offers a practical means of assessing both the effectiveness of the compression and the sensitivity of the observation to each model parameter throughout the inference process.

We found that mapping model parameters to PCA components is impractical, as retaining sufficient spectral variance requires too many components by the second round of inference. Similarly, the mapping between model parameters and hand-crafted spectral summaries proved less effective than with auto-encoder-based summaries, particularly for parameters associated with the narrow spectral line, as expected. This highlights the advantage of using learned, non-linear representations from auto-encoders over traditional dimensionality reduction or manually designed features.

\begin{figure}[!h]
    \centering
    \includegraphics[width=\linewidth]{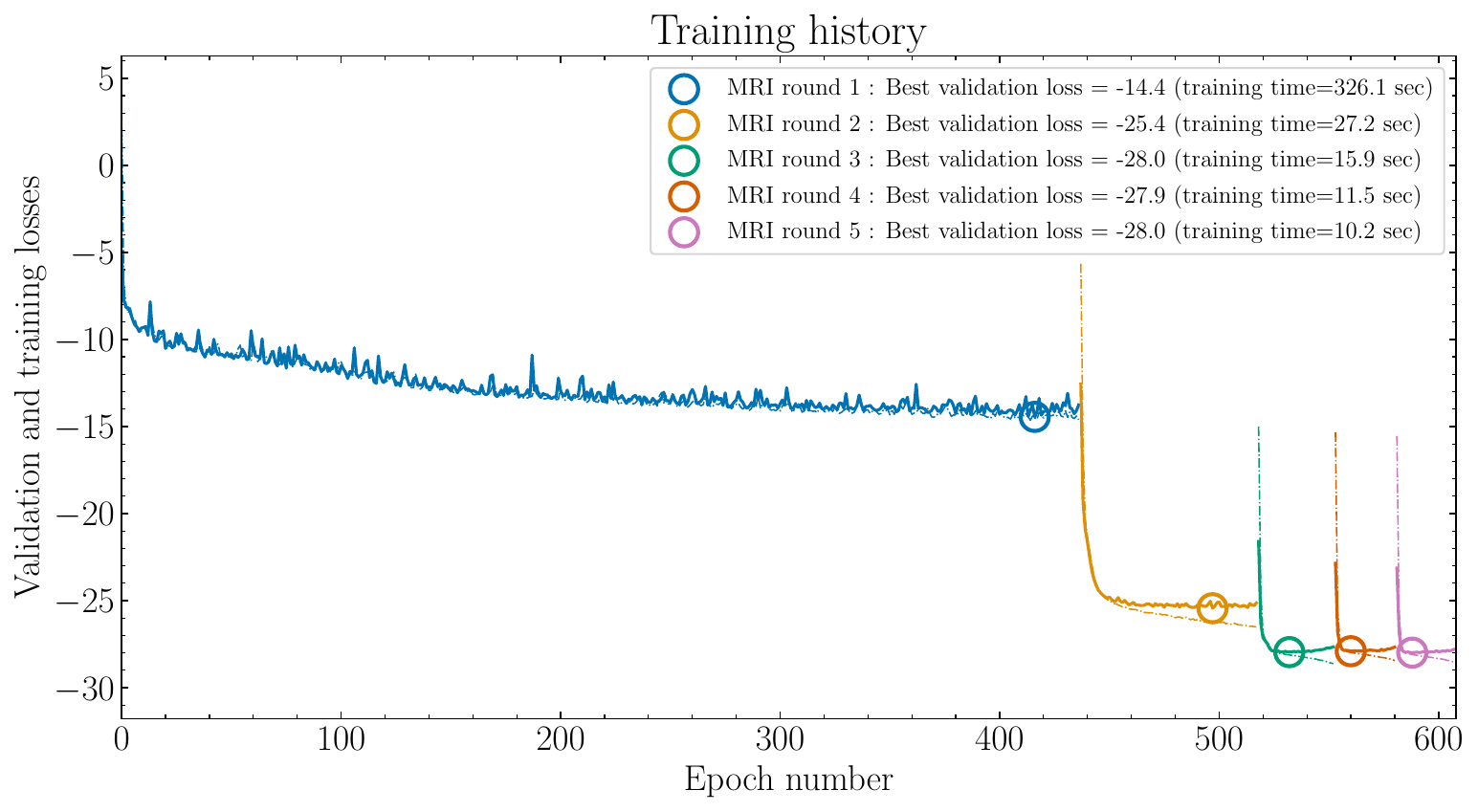}
    \caption{The validation loss (solid line) and training loss (dash-dot line) recorded during the NDE training, as a function of the cumulative epoch number for spectra compressed to 64 dimensions by the auto-encoder. The mathematical formulation of the loss function can be found in \cite{greenberg2019automaticposteriortransformationlikelihoodfree, deistler2022truncated}. The flatness of the curves towards the end of the rounds indicates no significant overfitting and successful training of the neural density estimator. Overfitting would appear as a sharp increase in validation loss, diverging from the training loss, signaling that the model captures noise or overly complex patterns in the training data. The training time is indicated on the plot, being the largest in the first inference round.}
    \label{fig:test-case-I-training-history}
\end{figure}

Figure \ref{fig:test-case-I-training-history} presents the validation and training losses of the NDE training, across five inference rounds for spectra reduced to 64 dimensions using the auto-encoder. The plot shows no evidence of overfitting: the validation loss remains stable and does not increase with further training epochs. The training converges by the third round, with only marginal improvements in subsequent rounds. In contrast, using Principal Component Analysis (PCA) for dimensionality reduction leads to overfitting as early as the first round. Notably, constraining the number of spectral summaries to fewer than 100, such as by considering 10 adjacent energy intervals, effectively mitigates overfitting in these cases.

\begin{figure*}[!t]
    \centering
    \includegraphics[width=0.3\linewidth]{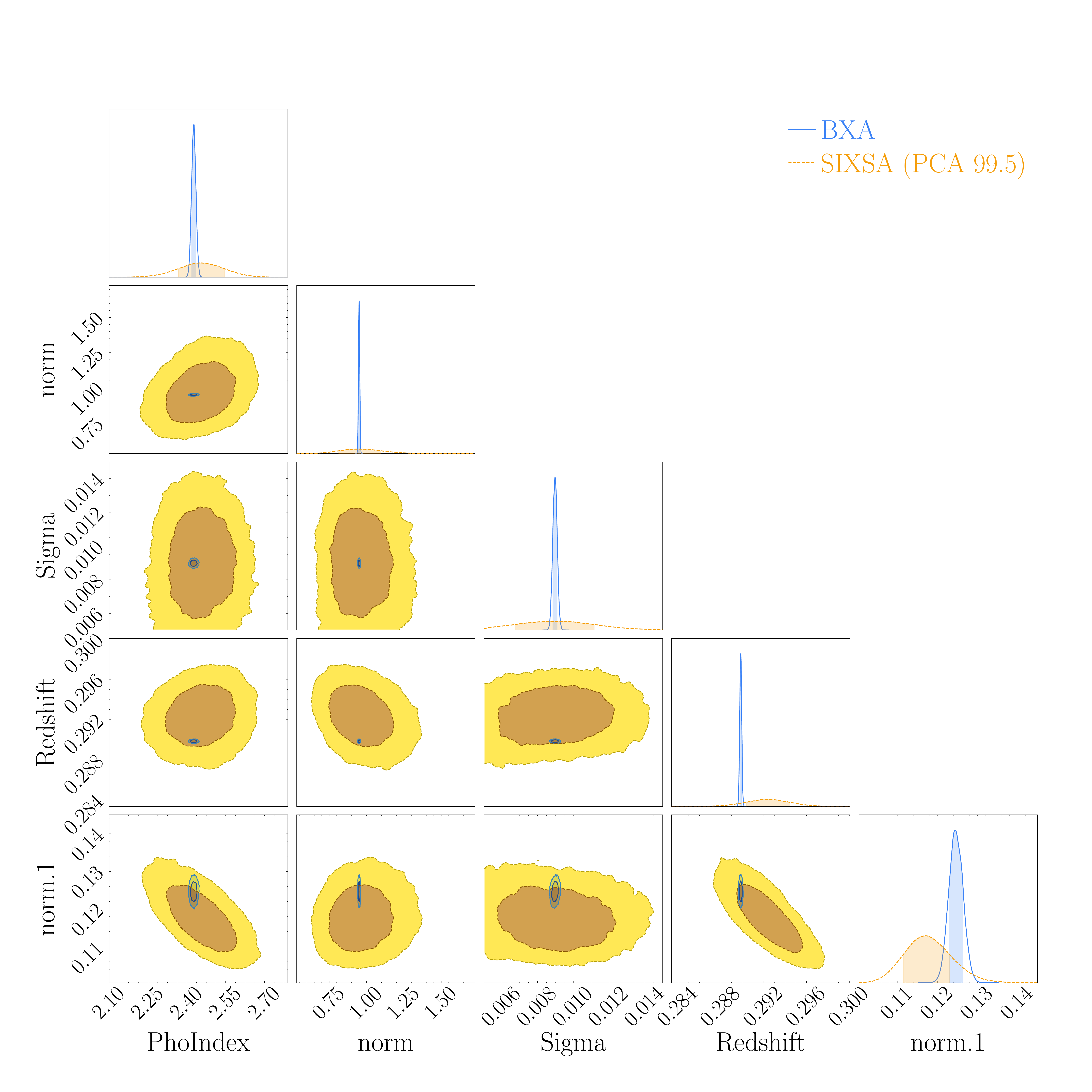}
    \includegraphics[width=0.3\linewidth]{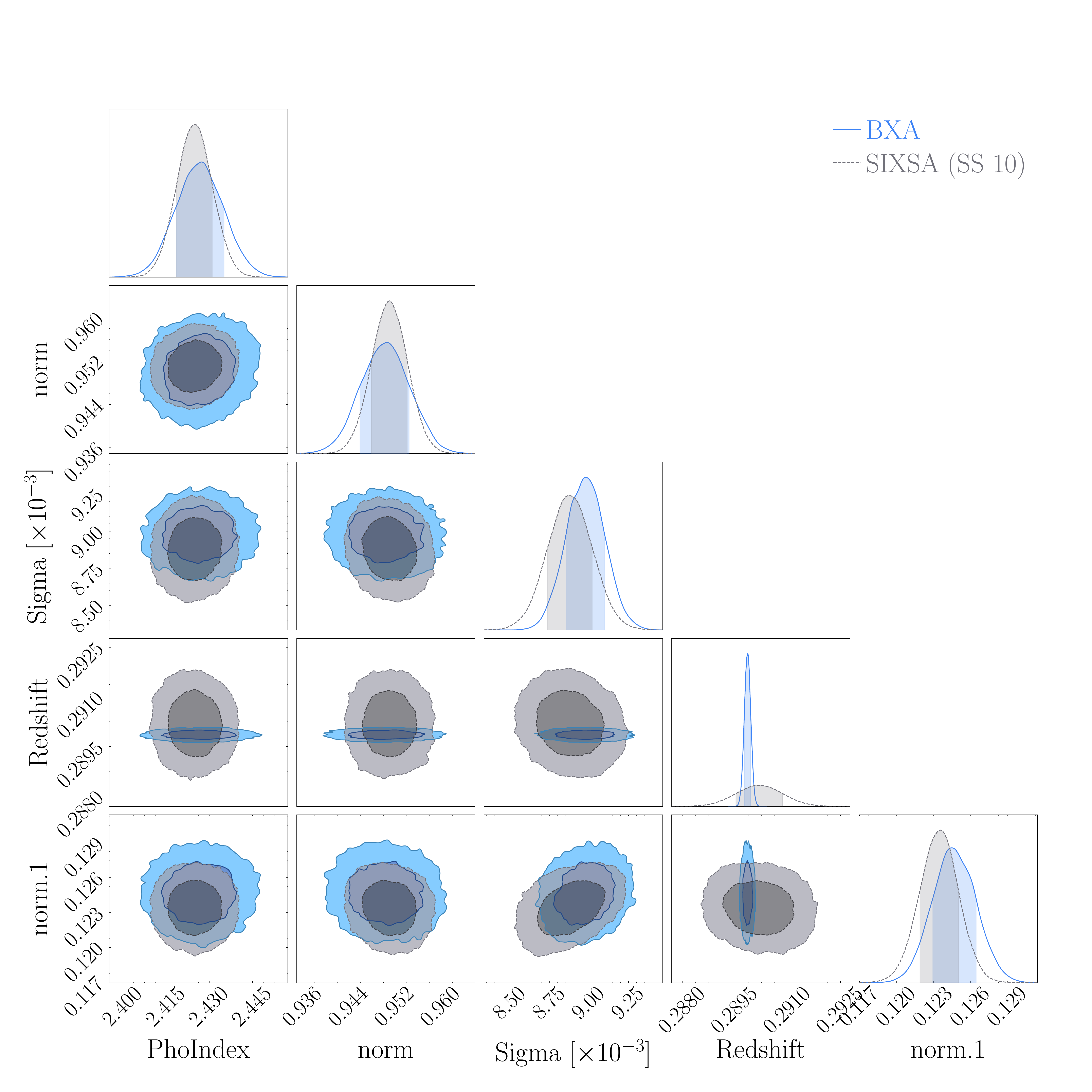}
    \includegraphics[width=0.3\linewidth]{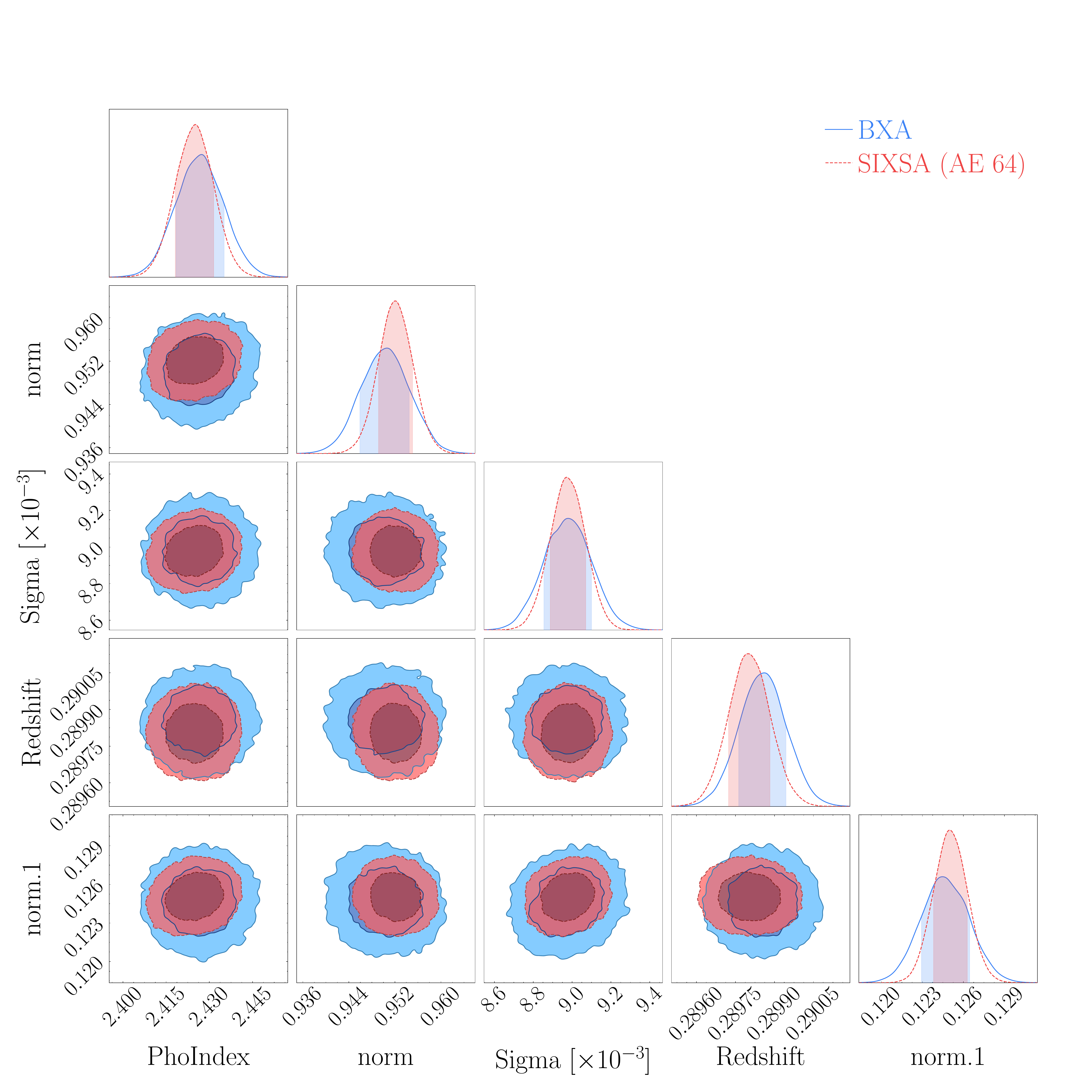}    \caption{Comparison of the posteriors for three dimension reduction techniques with respect to the reference posteriors computed by \bxa. Left: the PCA for which we retain 99.5\% of the sample variance at each round. Center: spectral summary statistics computed over 10 adjacent energy intervals. Right: our compact auto-encoder for which the spectra are reduced to a latent space of 64 dimensions. Over the 5 parameters of the model, the auto-encoder is clearly performing better than the two other reduction techniques.}
    \label{fig:comparison_of_three_dimension_reduction_schemes}
\end{figure*}

We now compare the posteriors obtained using the three dimensionality reduction techniques implemented in this work, using the \bxa\ posteriors as a reference. The results are presented in Figure \ref{fig:comparison_of_three_dimension_reduction_schemes}. For the five model parameters, the PCA shows the weakest performance. This is due to its limited ability to effectively reduce the dimensionality of the training spectra as they approach the target observation, which results in overfitting. Spectral summaries  offer some improvements, particularly for the power law parameters, but fail to recover the posterior for the line redshift, as expected. In contrast, the auto-encoder consistently performs better than both the PCA and summary statistics, producing posteriors that more closely align with those from \bxa. However, some differences remain, which will be corrected with importance sampling.
 
\begin{figure}
    \centering
    \includegraphics[width=0.85\linewidth]{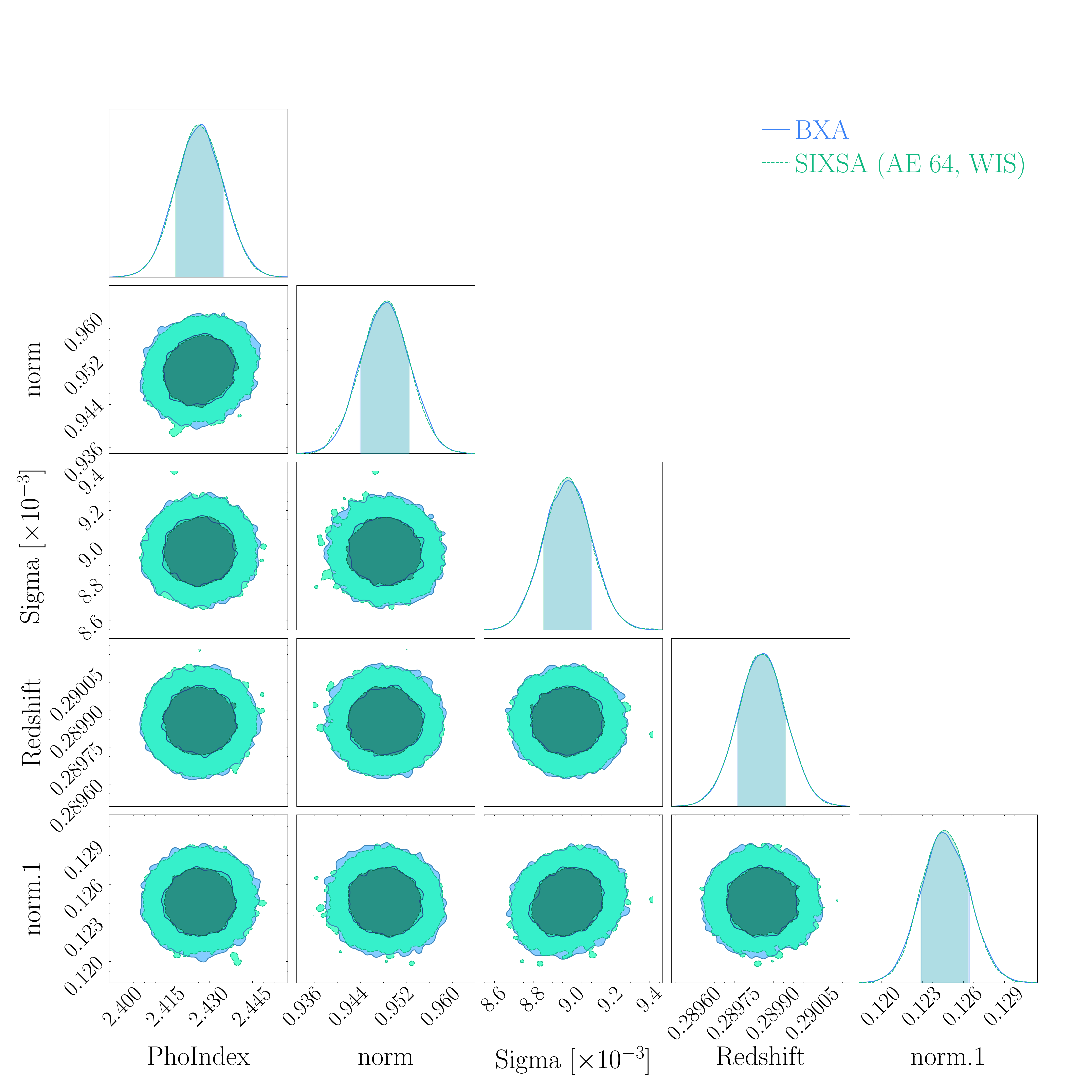}
    \caption{Comparison between the reference \bxa\ posteriors with the \sixsa\ posteriors (see Figure \ref{fig:comparison_of_three_dimension_reduction_schemes}, right panel) corrected by weighted importance sampling. The match is excellent for all 5 model parameters.}
    \label{fig:BXA_versus_SIXSA_corrected}
\end{figure}

 We thus apply importance sampling to correct the approximate posteriors, using 200,000 samples to compute the importance weights. The \sixsa\ corrected posteriors are shown in Figure \ref{fig:BXA_versus_SIXSA_corrected}, demonstrating an excellent match with the \bxa-derived posteriors across all five parameters. We have also corrected the approximate posteriors obtained with the PCA-based dimensionality reduction and the spectral summaries. It is interesting to note that the approximate posteriors obtained using spectral summary statistics can be effectively corrected. In particular, the redshift of the line, which was previously poorly recovered, aligns well with the reference derived from \bxa. In contrast, the posteriors derived from PCA-based dimensionality reduction deviate too significantly from the true distributions and cannot be reliably corrected using importance sampling.
 
\begin{figure}
    \centering
    \includegraphics[width=.995\linewidth]{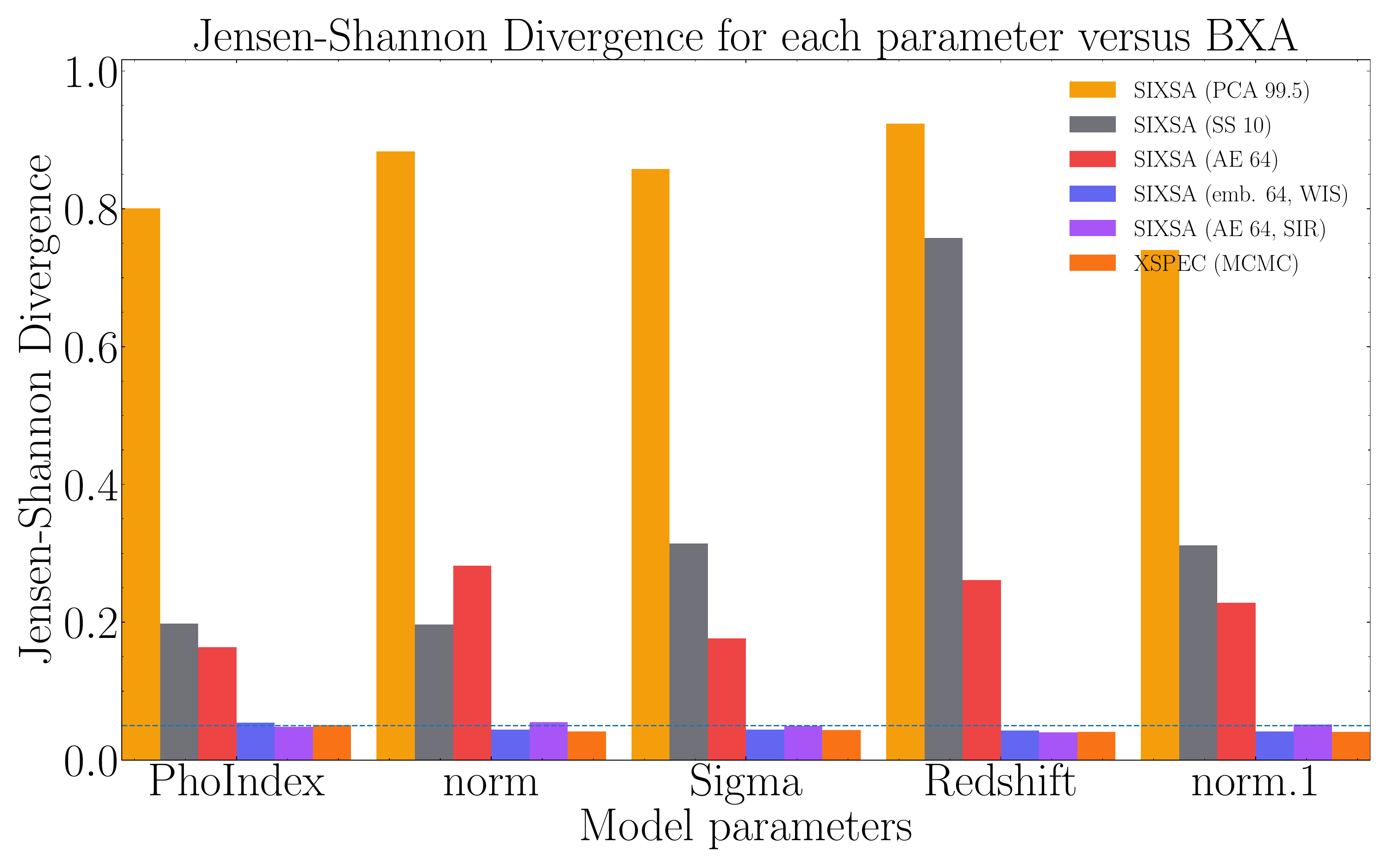}
    \caption{We show the JSD for each model parameter with respect to the reference posteriors obtained using \bxa. The JSD quantifies the similarity between posterior distributions; a smaller JSD indicates greater similarity. Below the dashed line, the posteriors are similar. We present the JSD values for three different dimension reduction techniques: PCA, spectral summaries, and the auto-encoder. Among these methods, the auto-encoder yields posteriors most similar to those from \bxa. We also consider a case in which an embedding network, identical to the encoder component of the auto-encoder, is used to compress the spectra before passing them to the neural density estimator. After applying importance sampling, the resulting posteriors closely match those from \bxa. We also show the \sixsa\ posteriors corrected with the sampling importance sampling technique. They are also very close to the \bxa\ reference posteriors. Finally, for the sake of the comparison, we include the posteriors computed using the \xspec\ MCMC method.}
    \label{fig:JSD}
\end{figure}

A quantitative way to assess the improvement of the posteriors relative to the \bxa\ reference is by computing the Jensen–Shannon Divergence (JSD), which measures the similarity between two probability distributions \citep{LinIEEE}. A smaller JSD value indicates a higher similarity. As shown in Figure \ref{fig:JSD}, the posteriors corrected by importance sampling exhibit the lowest JSD across all five model parameters, confirming their close agreement with the \bxa\ posteriors. Figure \ref{fig:JSD} extends the comparison to the approximate posteriors of Figure \ref{fig:comparison_of_three_dimension_reduction_schemes}, indicating that the posteriors obtained with \sixsa\ without importance sampling correction are the closest to the \bxa\ ones. The same figures shows that the \sixsa\ posteriors corrected with the sampling-importance-resampling method are also consistent with the \bxa\ ones.

It is entirely feasible to use the encoder component of the auto-encoder as an embedding network—that is, a neural network that takes the spectra as input, learns summary statistics, and passes these to the neural density estimator. In this configuration, the encoder parameters are trained jointly with those of the density estimator. However, this joint optimization requires a significantly larger simulation budget compared to pretraining the auto-encoder outside the inference loop. In our experiments, increasing the training sample size from 10,000 to 50,000 and applying importance sampling yielded posteriors that closely match those obtained with \bxa, as shown in Figure \ref{fig:JSD}. Despite this, the training history indicated signs of overfitting, suggesting that even more training samples may be necessary for stable convergence. From both a computational and practical standpoint, we found that training the auto-encoder separately, ahead of the inference process, offers greater stability and efficiency.  

\subsection{Test case II : \texttt{relxillp} and the role of importance sampling}
  
Increasing the complexity of the model, we now consider the \texttt{relxillp3.4} model\footnote{The \texttt{RELXILL\_RENORMALIZE} environment variable was set to 1.}, considering seven free model parameters \citep{Garcia_2014ApJ...782...76G,Dauser_2016A&A...590A..76D}. The range of model parameters are listed in Table \ref{tab:parameter_limits_caseII}. This is a simplified version of the model of \cite{Barret_2019A&A...628A...5B}, who tested the sensitivity of \xifu\ to supermassive black hole spins, yet without deriving the full posteriors of the model parameters. We thus simulate spectra with moderate statistics (1 million counts), corresponding to an exposure time of a 1 mCrab source ($\sim 2 \times 10^{-11}$ ergs/cm$^2$/s, 2-10 keV) for $\sim 10 $ kilo-seconds. The scope here is to highlight the role of importance sampling.
\input{figures/caseII/parameter_limits}
For this model, we again perform five rounds of inference, each using a truncated proposal and 20,000 simulated spectra. Despite the increased model complexity, we retain a latent space dimension of 64 for the auto-encoder. To apply importance sampling corrections, we replace the exact likelihood computation with the \likelihoodemulator\ trained on 50,000 exact likelihoods computed from the posterior samples obtained at the fifth inference round. The \likelihoodemulator\ performance is evaluated by comparing predicted versus actual Log-likelihood values for a test set of model parameters not seen during the training.

\begin{figure}
    \centering
    \includegraphics[width=0.995\linewidth]{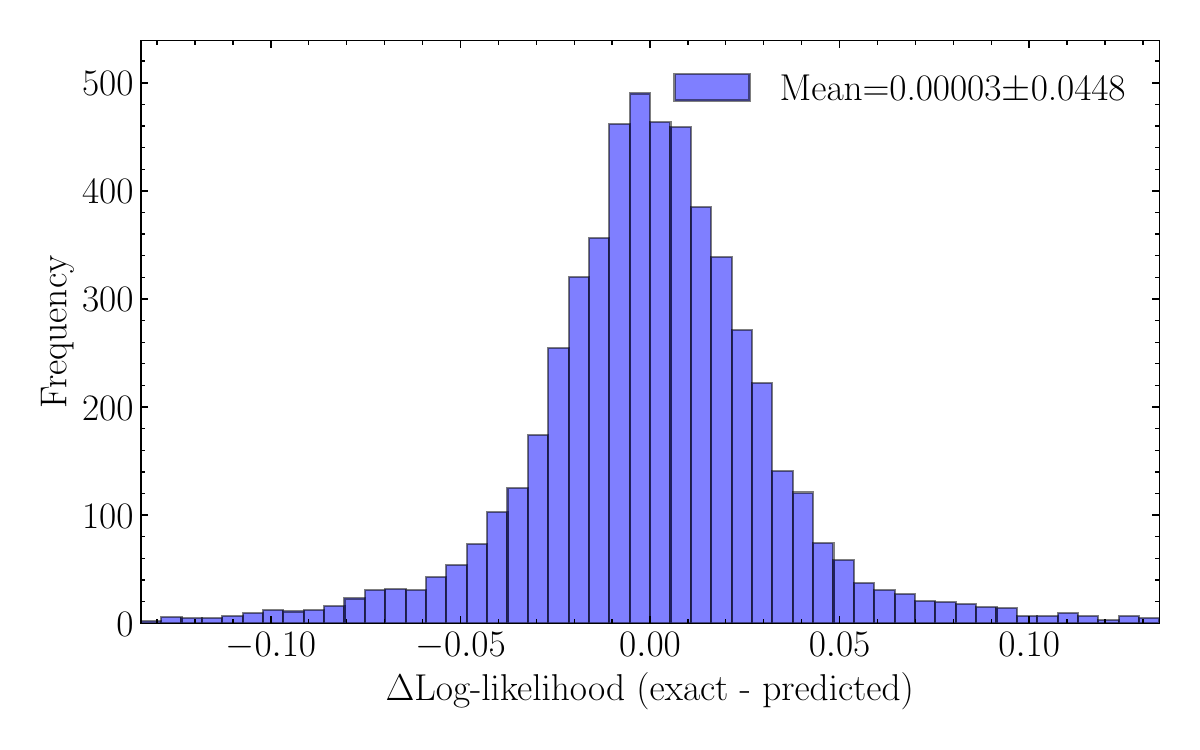}
    \caption{Histogram of the prediction errors of the \likelihoodemulator for the \texttt{relxillp} model (with the mean and the standard deviation of the error). From 50,000 drawing posterior samples derived from the fifth inference round, 50,000 exact likelihoods were used for training the neural network. 5,000 additional likelihood samples were used for the test set and compared with the predictions from the \likelihoodemulator. The accuracy reached is sufficient for performing accurately the importance sampling correction.}
    \label{fig:cstat_learner}
\end{figure}

We then compare the results obtained using the two importance sampling approaches implemented in \texttt{sbi}: standard weighted importance sampling and sampling-importance-resampling (SIR, the second method described in \S\ref{sir}). In the SIR case, generating 50,000 posterior samples with an oversampling factor of 32 requires $32 \times 50,000 = 1.6$ million likelihood evaluations, which without the \likelihoodemulator\ would translate to an equal number of costly simulations. In such cases, importance sampling can become the dominant component of the simulation budget. For comparison, computing all 1.6 million exact likelihoods takes roughly one hour using 16 cores in parallel, whereas training the \likelihoodemulator\ takes only a few minutes, and evaluating 1.6 million approximate likelihood functions requires about 20 seconds. As in previous experiments, we take the \bxa\ posteriors as the reference, and we additionally include posterior estimates derived from an \xspec\ MCMC run. In Figure \ref{fig:cstat_learner}, we show the performance of the \likelihoodemulator\ on a test set of 5000 model parameters, from the posteriors computed at the fifth round of inference. The standard deviation of the error between the predicted and exact log-likelihood is less than $0.05$. We also verified that importance sampling remains effective even with a less accurate \likelihoodemulator. For instance, training the \likelihoodemulator\ on a smaller dataset of 20,000 samples instead of 50,000 still yields acceptable posterior corrections, indicating that importance sampling can tolerate a moderate level of error in the approximation of the likelihood function.

\begin{figure}
    \centering
    \includegraphics[width=\linewidth]{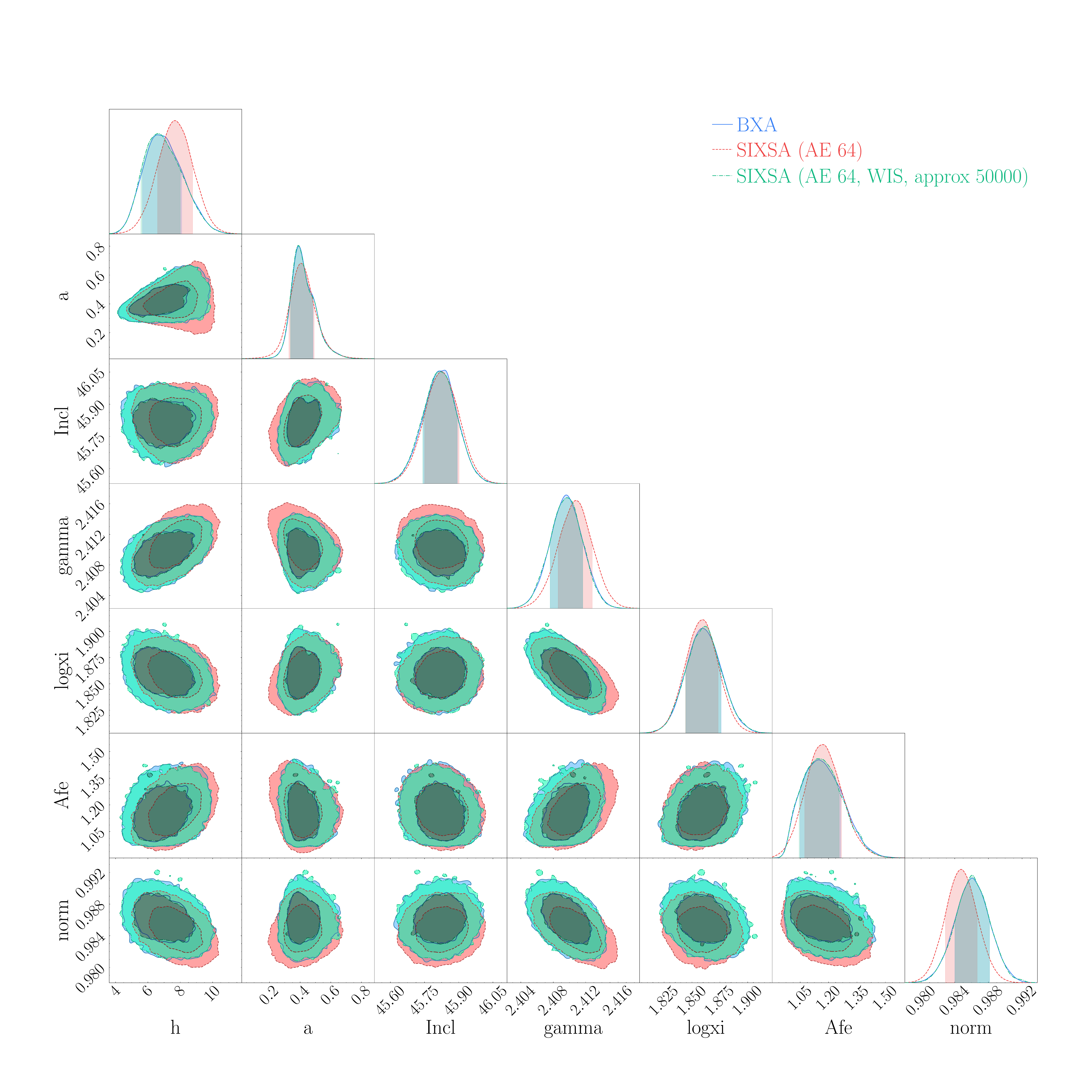}
    \caption{Posteriors from \bxa, the fifth inference round before and after weighted importance sampling, when the likelihood is computed with the \likelihoodemulator. The \sixsa\ corrected posteriors and the \bxa\ ones are undistinguishable.}
\label{fig:relxill_importance_sampling}
\end{figure}

In Figure \ref{fig:relxill_importance_sampling}, we compare the posterior distributions with the \bxa\ posteriors, both before and after applying weighted importance sampling. Several key observations can be made. First, within the simulation budget used for the five inference rounds (i.e., five sets of 20,000 simulations), the normalizing flows tend to produce Gaussian-like posterior shapes. Second, importance sampling effectively corrects these approximations, yielding posteriors that closely match those obtained with \bxa. This correction recovers the more complex posterior structures, particularly for the corona height, black hole spin, and iron abundance. The behavior of these parameters is consistent with the mapping between the model parameters and the latent space, which indicated that these three parameters were less well-constrained compared to the others (see Appendix \ref{Appendix A} for a more extreme example of this effect).

\begin{figure}
    \centering
    \includegraphics[width=\linewidth]{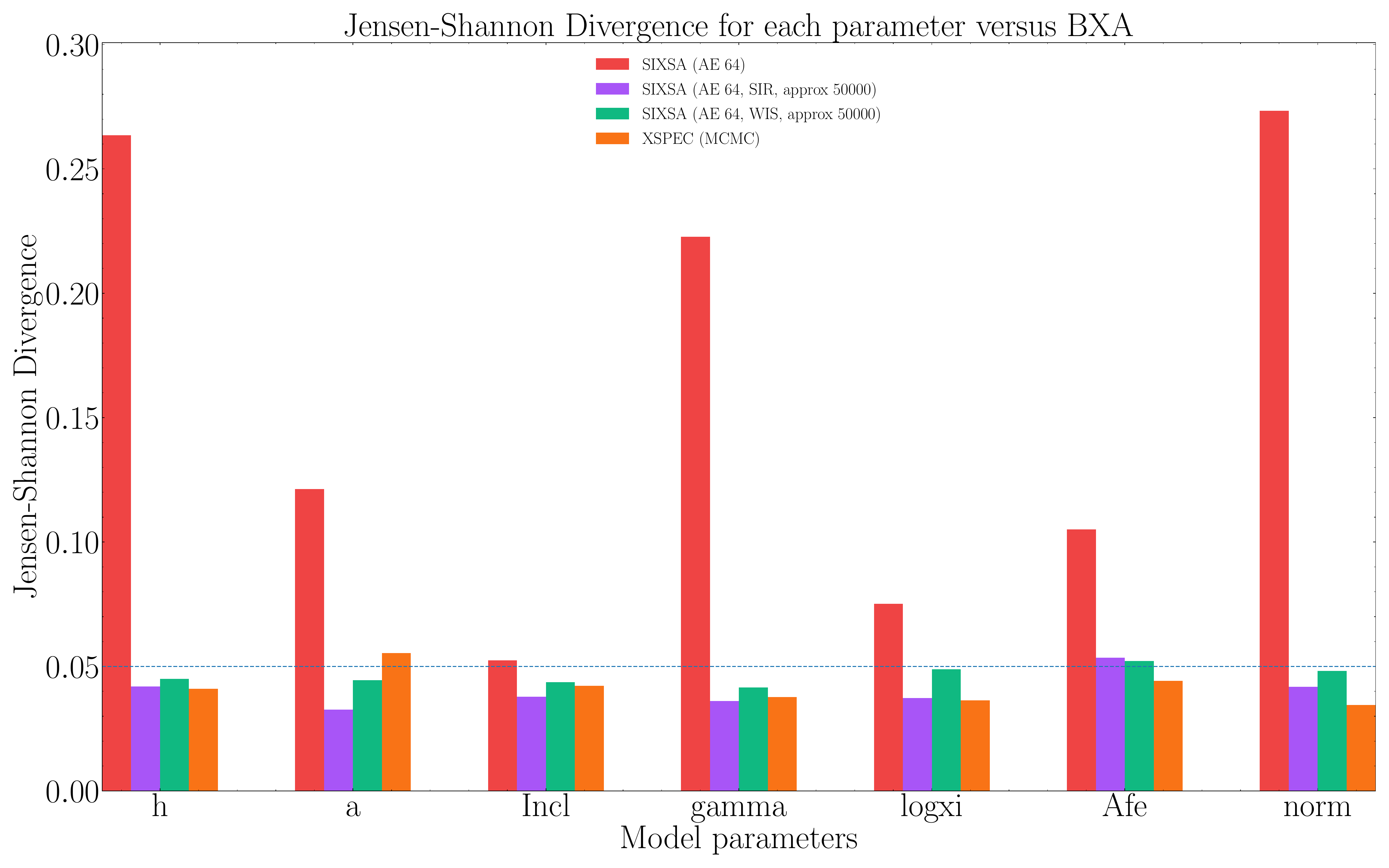}
    \caption{Comparison between the posteriors before and after importance sampling correction with the reference \bxa\ posteriors: weighted importance sampling and sampling-importance-resampling, considering the \likelihoodemulator\ for computing the likelihood, trained and tested on 50,000 samples. We also show the \xspec\ MCMC posteriors, fully comparable to the \sixsa\ ones after importance sampling. This figure highlights the role of importance sampling. }
    \label{fig:relxill_jsd}
\end{figure}

In Figure \ref{fig:relxill_jsd}, we present the JSD for the two importance sampling methods, training the \likelihoodemulator\ on 50,000 samples. For comparison, we also include the \xspec\ MCMC posteriors. The results demonstrate that both importance sampling methods implemented in \sbi\ yield consistent corrections. Notably, for that particular case, no bias is observed in the sampling-importance-resampling method, even when using approximate likelihood estimates. 

\begin{figure*}[!t]
    \centering
    \includegraphics[width=0.575\linewidth]{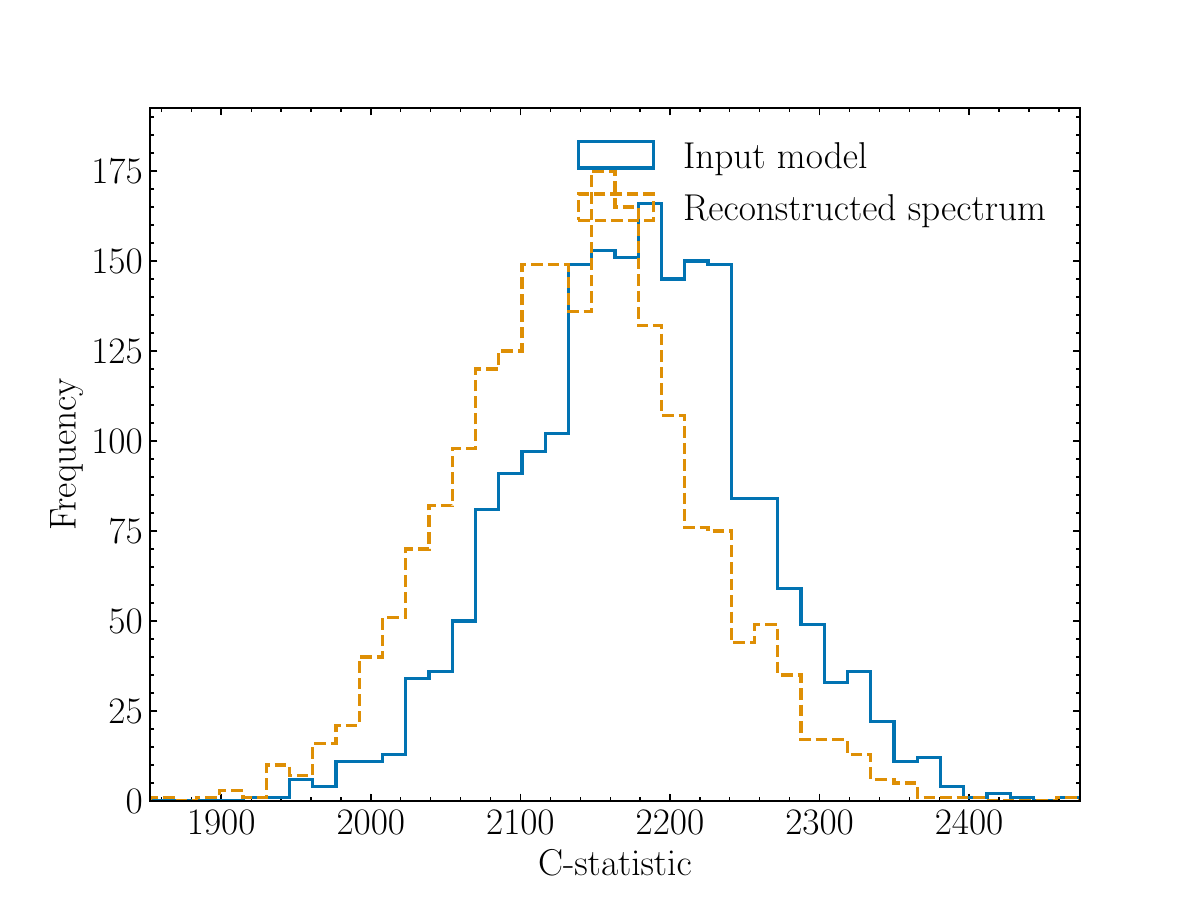}    \includegraphics[width=0.325\linewidth]{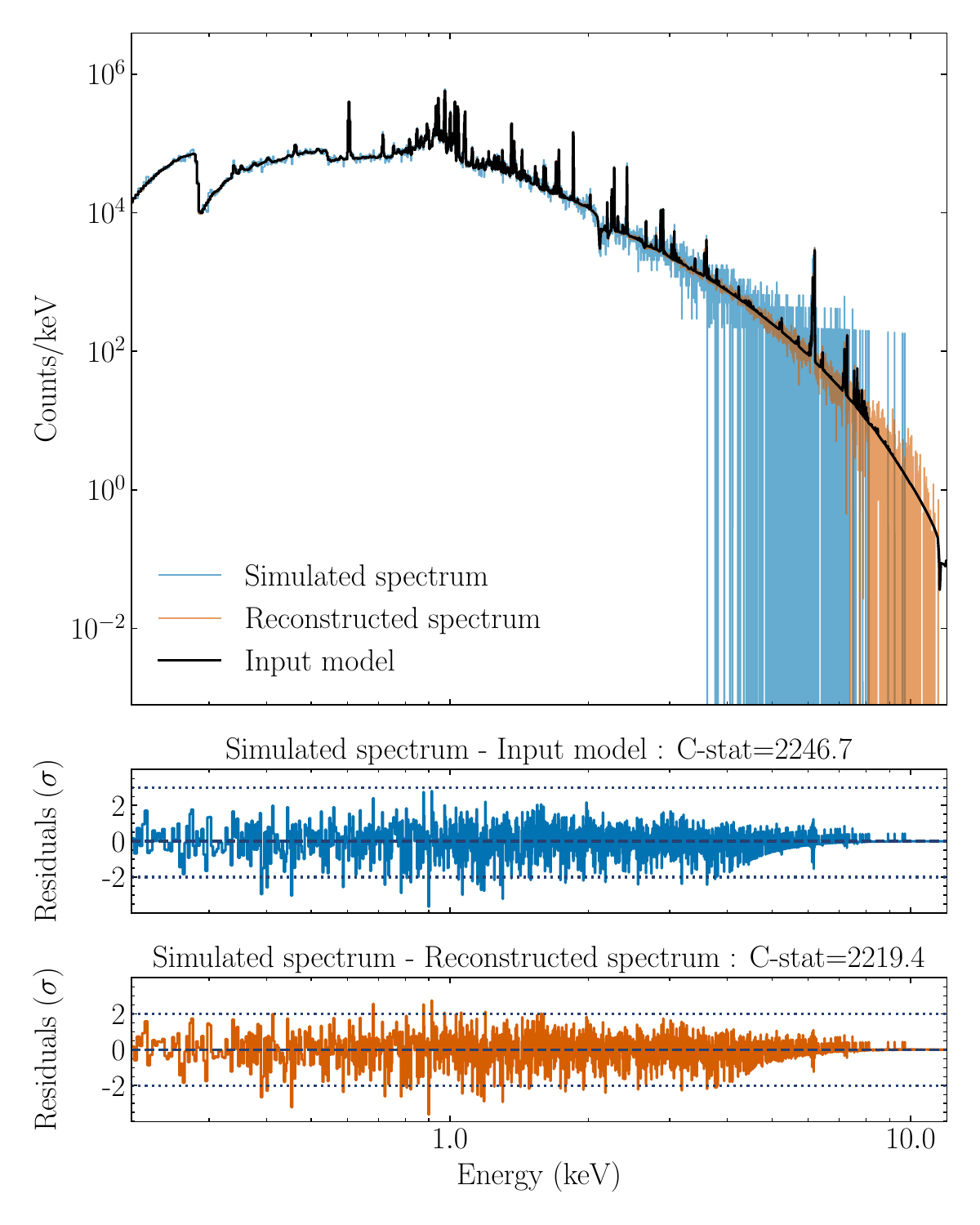}
    \caption{Left: Histogram of \cstat\ of a sample of 2000 simulated spectra, with model parameters drawn from the truncated proposal after the first round of inference. The \cstat\ are computed with respect to the input model (blue solid line) and the reconstructed spectrum (dashed orange line), without any minimization.  Right: A random example of such a sample spectrum (blue), with its input model (black solid line) and the reconstructed spectrum, from the decoder part of the auto-encoder (in orange). The residuals, expressed in $\sigma$, between the simulated spectrum and both the input model and the reconstructed spectrum are shown in the bottom panels, with the same colors. The corresponding \cstat\ (without minimization) is listed for indication at the top of each sub-panel. }
    \label{fig:2bvapec_cstat_comparison}
\end{figure*}
\subsection{Test case III : a double \texttt{bvapec} model}
The models discussed above were selected to showcase the auto-encoder ability to retain narrow line features superimposed on a broad continuum, as well as to illustrate the benefits of importance sampling. However, these models do not fully capture the complexity typical of \xifu-like spectra. To address this, we consider a more realistic model composed of the sum of two \texttt{bvapec} components, both sharing the same redshift. The elemental abundances are fixed to their default values, and the velocity for both components is set to 100 km/s. This choice is motivated by the fact that, in the statistical regime we aim to explore, the velocity cannot be meaningfully constrained (see Table \ref{tab:parameter_limits_caseIII} for the range of variation of the free model parameters).
\input{figures/caseIII/parameter_limits}
This model introduces significant degeneracy and is expected to produce broad or multi-modal posterior distributions, posing a challenge for traditional fitting methods. To accommodate this added complexity, we consider a training sample size of 50,000 and keep the latent space dimension of the auto-encoder to 64. We find that a dimension of 128 performs equally well, but that increasing it beyond 128 offers no additional benefits in general.

In Figure \ref{fig:2bvapec_cstat_comparison}, we compare the histogram of the \cstat\ values for a set of simulated spectra, with model parameters sampled from the truncated proposal distribution after the first round of inference. The \cstat\ values are computed with respect to both the input models and the reconstructed spectra. An illustrative example of a simulated spectrum along with its corresponding input model and reconstructed spectrum is also included in the figure. As shown, the two \cstat\ histograms are aligned, indicating that the auto-encoder has successfully recovered the overall spectral shape across the dataset. Additionally, the model accurately reproduces narrow spectral line features, demonstrating its effectiveness in capturing both global and localized spectral features. Looking at the latent space distributions reveals that they are well-regularized, with most dimension histograms exhibiting approximately Gaussian-like shapes.

The entire pipeline including the simulations, the auto-encoder training, neural density estimator training, truncated proposal generation, and importance sampling correction completes in slightly more than 1 hour with the bulk of the run time coming from the first training round  of the auto-encoder. For comparison, we compute reference posteriors using \bxa\ on the original prior with 3,000 live samples. Running \bxa\ in parallel with \texttt{mpiexec} on 16 cores requires approximately 14 hours and over 10 million likelihood evaluations.

\begin{figure}[!t]
    \centering
    \includegraphics[width=\linewidth]{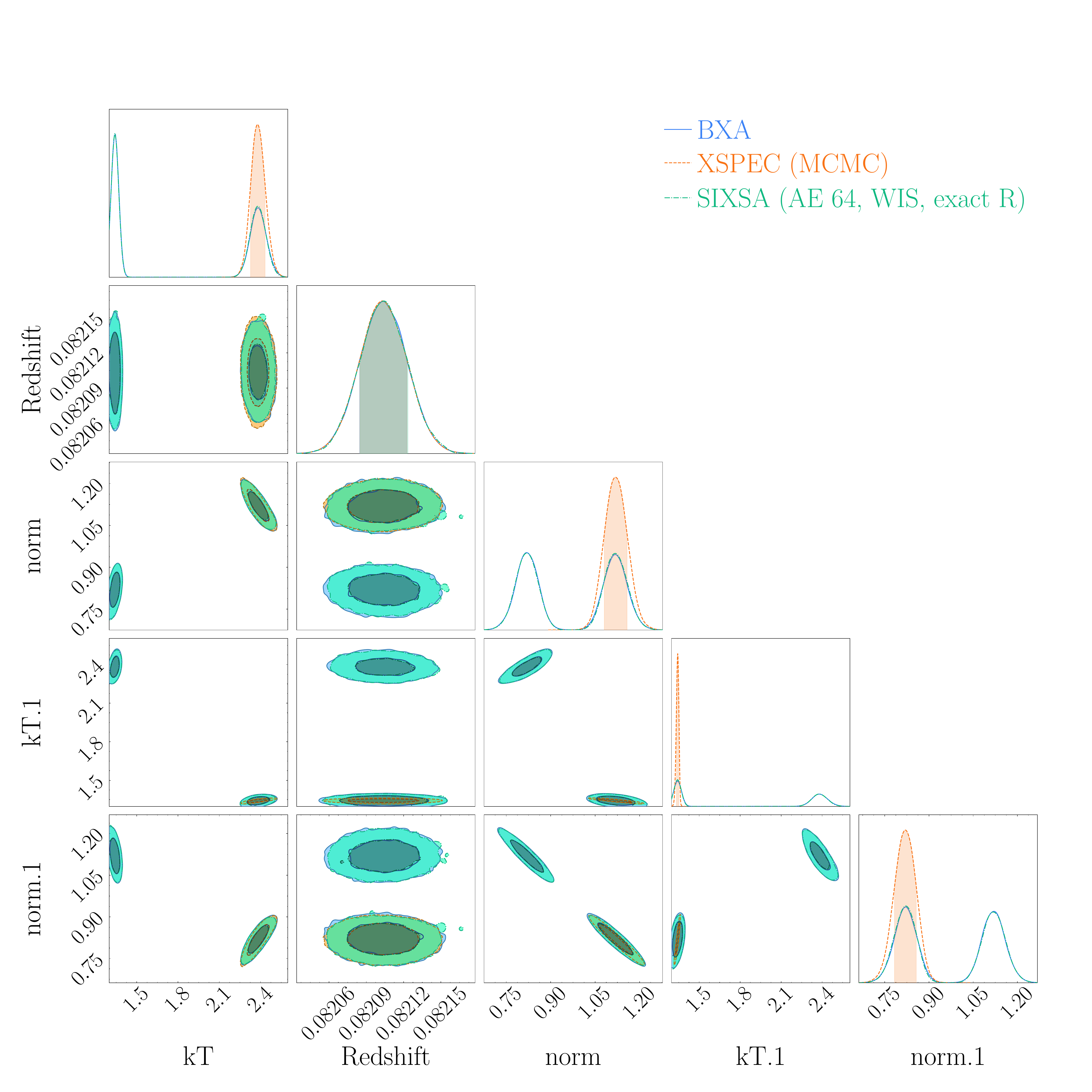}
    \caption{Comparison between the \bxa\ reference posteriors with the \xspec\ MCMC posteriors and the importance sampling corrected \sixsa\ posteriors. The match between \sixsa\ and \bxa\ is excellent, and \sixsa\ runs 20 times faster than \bxa\ on such a case. MCMC fails to capture the bimodal posterior distribution.}
    \label{fig:2bvapec_bxa_sixsa_corrected}
\end{figure}
In Figure \ref{fig:2bvapec_bxa_sixsa_corrected}, we compare the \bxa\ posteriors with the importance sampling–corrected \sixsa\ posteriors, demonstrating that the results are in strong agreement. For additional context, we include posteriors obtained from a standard \xspec\ MCMC run. As expected, the MCMC approach struggles to adequately sample the full parameter space in the presence of a highly multi-modal posterior distribution. 

\begin{figure}
    \centering
    \includegraphics[width=\linewidth]{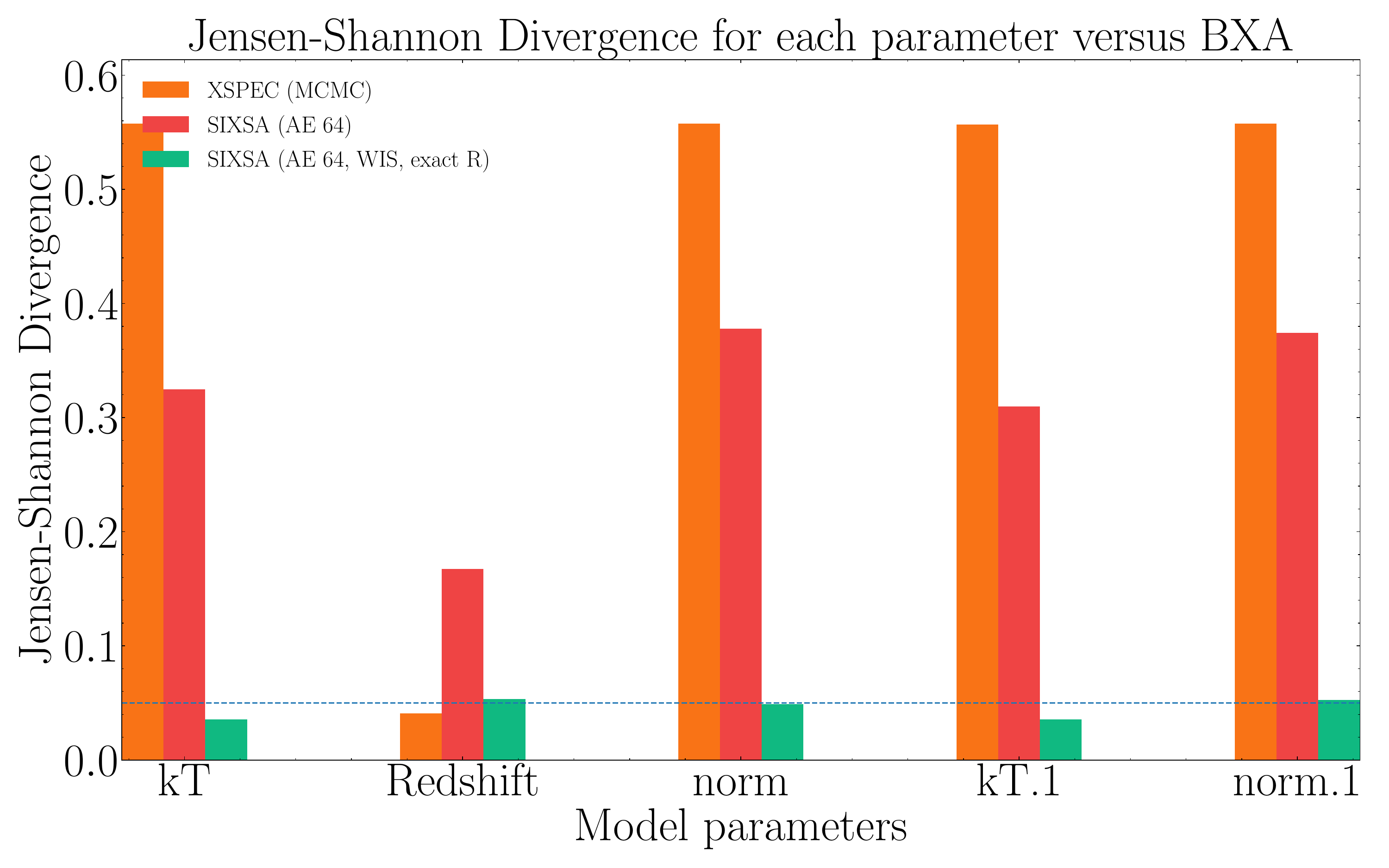}
    \caption{JSD between the posteriors derived from an \xspec\ MCMC run (blue), the ones derived by \sixsa\ at the fifth round of inference prior (orange) and after the importance sampling correction (green). }
    \label{fig:2bvapec_jsd}
\end{figure}

\begin{figure}
    \centering
    \includegraphics[width=\linewidth]{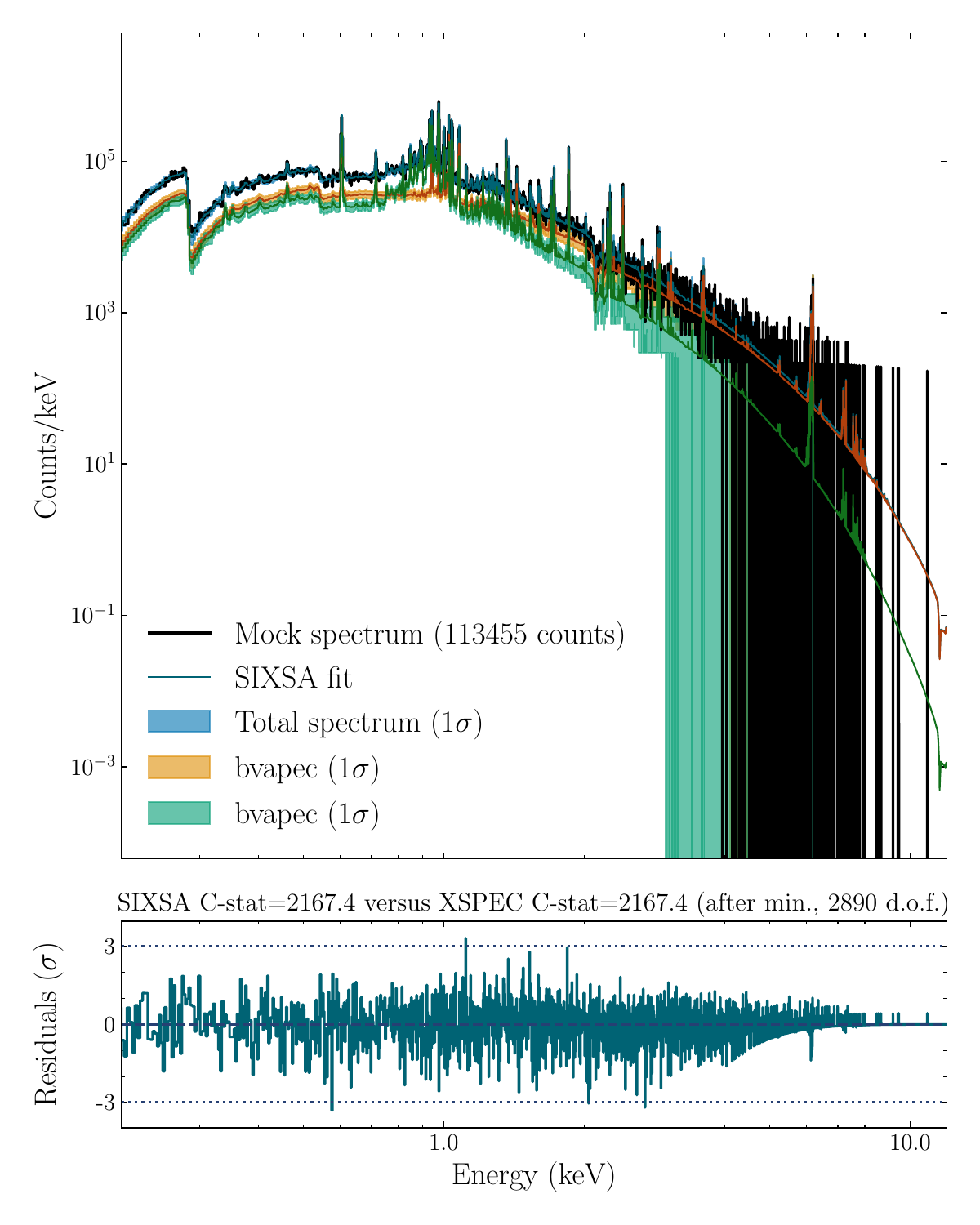}
    \caption{Posterior predictions for the double \texttt{bvapec} model. In order to separate the posterior samples for the two model components, a gaussian mixture model was fitted to the distribution of the normalization of the first \texttt{bvapec} component. The posterior samples could then be labeled and separated. The \cstat\ associated with the median of 1000 drawn posterior samples is equal to the \cstat\ computed by \xspec\ after minimization.}
    \label{fig:2bvapec_folded}
\end{figure}

In Figure \ref{fig:2bvapec_jsd}, we present the JSD between the \bxa\ posteriors and three alternative posterior estimates: those obtained from \xspec\ MCMC, the approximate posteriors produced by \sixsa\ at the fifth inference round, and the corresponding posteriors after importance sampling correction. As shown, even the uncorrected \sixsa\ posteriors are closer to the \bxa\ reference than those from \xspec\ MCMC, particularly in capturing the bimodal nature of the posterior distributions for the four degenerate model parameters. This demonstrates that the combination of dimensionality reduction via auto-encoders, iterative inference, and importance sampling enables accurate recovery of complex posterior structures, even in highly degenerate cases.
In Figure \ref{fig:2bvapec_folded}, we present the best-fit \sixsa\ model for the double \texttt{bvapec} case, along with posterior predictive spectra for each of the two components. To disentangle the contributions from the overlapping components, a two-component Gaussian Mixture Model was applied to the posterior samples, splitting them based on the normalization parameter of the first \texttt{bvapec} component. Despite the spectral proximity of the two components, \sixsa\ successfully reconstructs them with high fidelity. Notably, the median of the posterior samples yields a \cstat\ value identical to the minimum \cstat\ obtained via \xspec\ minimization, further validating the accuracy of the inferred model.

\begin{figure*}
    \centering
   \includegraphics[width=\linewidth]{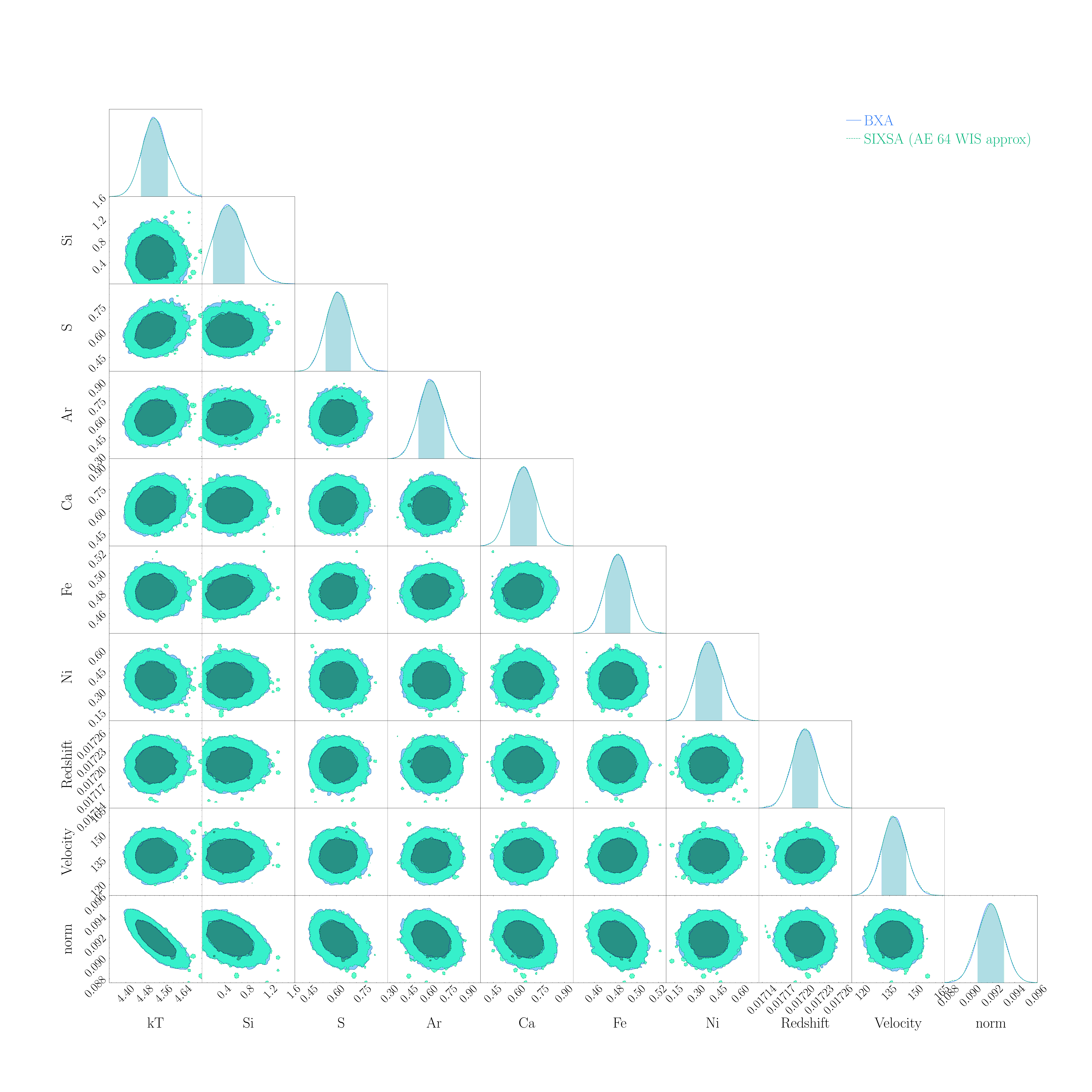}    
         \caption{Comparison between the \sixsa\ and \bxa\ posteriors for a \xrism-Resolve observation of the Perseus galaxy cluster. The \sixsa\ ones have been corrected by importance sampling, using likelihoods approximated with the \likelihoodemulator. As can be seen there is an excellent match between the two. More sophisticated analysis, not in the scope of this paper, is required to derive meaningful abundances from this observation.}
    \label{fig:xrism_posteriors}
\end{figure*}

\subsection{Application to \xrism-Resolve data} While mock spectra generated by the same simulator as the one generating the training set are useful for benchmarking, the true test of any deep learning technique lies in its application to real observational data. We now consider, as the ultimate challenge, tests with data provided by the X-Ray Imaging and Spectroscopy Mission (\xrism)-Resolve high-resolution X-ray spectrometer, often considered as the precursor of \xifu\ \citep{Tashiro2025PASJ..tmp...28T,XRISM_first_light_paper2024PASJ...76.1186X,Ishisaki_2022SPIE12181E..1SI}. The microcalorimeter \textit{Resolve}, a 6~$\times$~6 pixel X-ray spectrometer, delivers exquisite data above 2 keV with a spectral resolution of 4.5 eV at 6 keV. For the sake of the exercise we consider an archival observation of the Perseus cluster (OBSID 000156000, so-called Perseus-C1 pointing, \cite{XRISM_perseus_2025arXiv250904421X}) taken on January 23rd, 2024 for an effective exposure time of $\sim 57$ ks. Standard screening criteria were applied to the observation. Only the highest resolution primary events events were retained. Optimal binning was applied to the extracted spectrum, computed as the sum of the spectra of all the good pixels, thus excluding pixels 12 and 27. In our analysis, we ignore the non X-ray background, and assume a single temperature \texttt{bvapec} model, leaving the abundance parameters of Si, S, Ar, Ca, Fe and Ni free (all other abundances are frozen to 0.7). We then keep the temperature, the redshift, the velocity free parameters of the fit. We consider data between 2 and 10 keV, including the region $6.567-6.620$ keV affected by resonant scattering (excluding it would lower the measured velocity). We warn the reader that the results derived below must be taken with cautions and shall not be used for scientific exploitation. More sophisticated analysis, including adding the non X-ray background in the analysis, considering multi-temperature models, spatial mixing\ldots is required to derive meaningful scientific information from this observation \citep{XRISM_Clusters_2025arXiv251006322X}: this is clearly outside the scope of this paper and left to the \xrism\ collaboration.

When dealing with real data, it is recommended to increase the number of simulations for training the neural density estimator. As for the double \texttt{bvapec} model above, we still consider 5 rounds of 50,000 spectra, with 100,000 in the first round. Given the size of the response file (20 times larger than the one used for \xifu), the simulations take more than 10 times longer than with \xifu, and become by far the dominant component of the run time. Ways to speed up the simulations are being investigated. We auto-encode the spectra again in 64 dimensions, and perform importance sampling with the \likelihoodemulator\ trained on 50,000 samples, and compute $32 \times 50,000$ likelihoods afterwards.  We run \bxa\ from the truncated prior proposal at the fourth round of inference. This reduces the number of likelihood estimates required by \bxa\ by more than one order of magnitude, making the \bxa\ run time comparable to the \sixsa\ run time.

Along the inference process, looking at the results of the \parameterretriever\ indicates that with this model, the observation is sensitive to all free parameters, yet with a reduced sensitivity to the Si abundance. The training process has converged after 5 rounds with no sign of overfitting. Figure \ref{fig:xrism_posteriors} compares the \bxa\ and \sixsa\ posteriors, indicating an excellent match between the two methods. We found that a setup with 25,000 simulations per round, and a latent space dimension of 32 would deliver comparable results. On the other hand, going to higher latent space dimension did not help, with clear signs of overfitting during the training of the neural density estimator. Both weighted importance and sampling-importance-resampling performed similarly.

\begin{figure}
    \centering
    \includegraphics[width=0.95\linewidth]{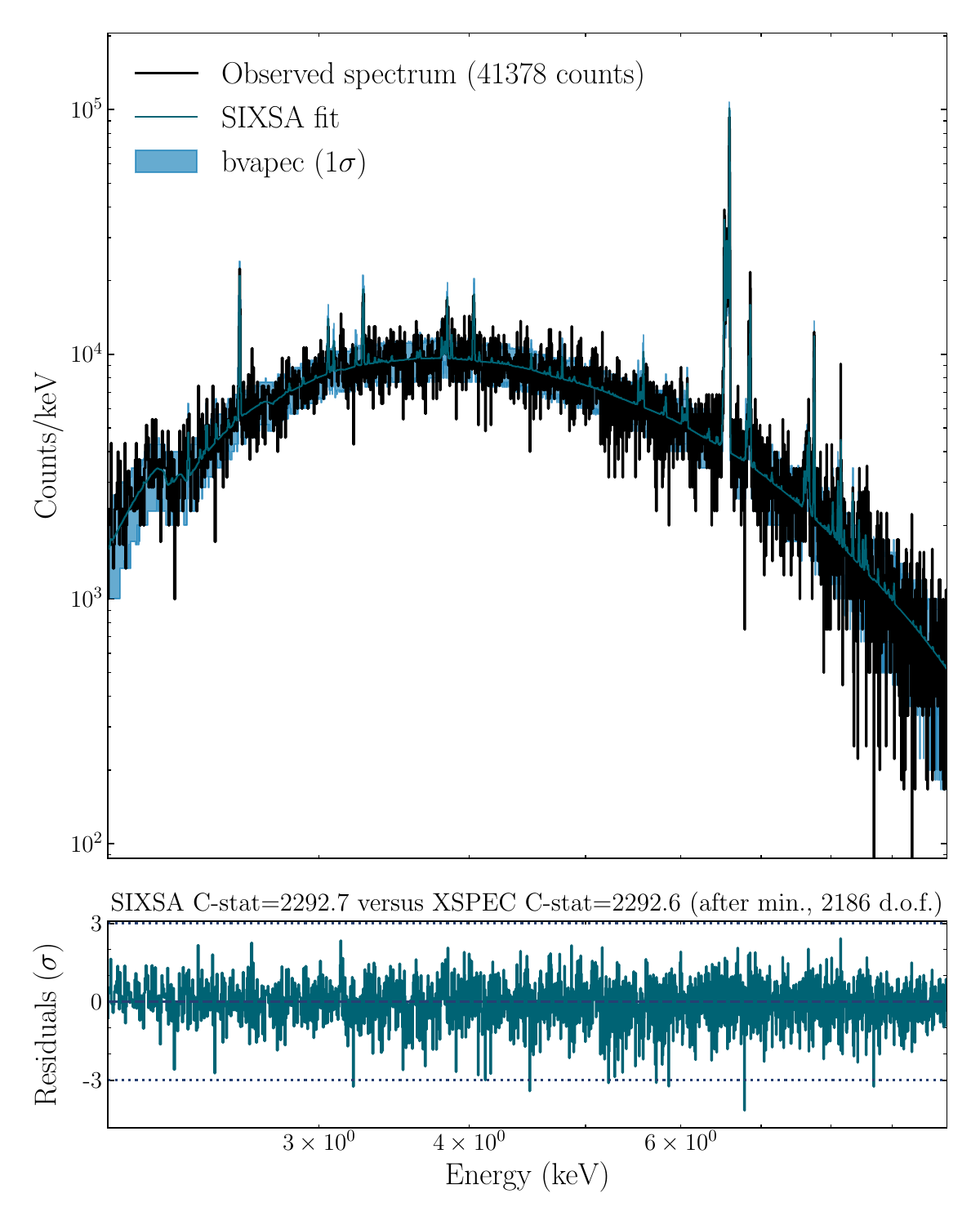}         \caption{The folded \xrism-Resolve spectrum of one snapshot observation of the Perseus cluster together with its reconstruction with a single temperature \texttt{bvapec} model. The \cstat\ associated with the median of 1000 drawn posterior samples is very close to the \cstat\ computed by \xspec\ after minimization.}
    \label{fig:xrism_folded_spectrum}
\end{figure}

Figure \ref{fig:xrism_folded_spectrum} shows the folded spectrum together with the posterior predictions. The \sixsa\ fit corresponds to the model whose parameters are the median of the posteriors. Its \cstat\ is comparable to the one of \xspec\ after minimization, indicating that for this model, \sixsa\ has found a solution very close to the best fit.
\section{Discussion}
\label{discussion}
With an auto-encoder used to reduce the dimension of \xifu-like spectra, we have shown that the training of the neural density estimator was stable and converged after a few rounds. Correcting the approximated neural posteriors, through importance sampling, we have shown that the \sixsa\ posteriors are identical to those derived by \bxa, even in the case of multi-modal posterior distributions. If run on the initial parameter space, \bxa\ takes 10-100 times more time than SIXSA, due to the larger number of likelihoods to be evaluated (10s of millions). Note, however, that to speed up \bxa, not excluding any regions of interest, one could reduce the parameter space to explore, either manually, or through the use of the truncated prior proposals derived by \sixsa, e.g. after the first inference round.

Auto-encoders nicely couple with multi-round inference because as the parameter space shrinks around the targeted observation, its performance increases, and the latent space becomes smooth and centered (almost Gaussian), easing the training of the neural density estimator. With the models considered here and our auto-encoder architecture, we have found that a latent space dimension below, say, 100 enables stable training and prevents overfitting. This is not the case with the PCA, which requires more and more components to retain the variance of the training sample as the parameter space shrinks. Similarly, our auto-encoder retains the information of narrow features in \xifu-like spectra, which is not the case with simple spectral summary statistics. Optimizing the auto-encoder architecture and its hyperparameter space is beyond the scope of the paper, but the potential of auto-encoders to reduce the dimension of X-ray spectra, coupled with a likelihood free inference followed by importance sampling to obtain accurate posteriors is the main result of this paper. In such a way, this concludes the initial studies presented in DB24 and BD25.

Although we have previously shown that obtaining comparable posteriors to the \bxa\ ones was doable without importance sampling, the possibility to correct the approximate posteriors derived by \sixsa, once the training has reached convergence, adds some degrees of freedom in its use. The performance of \sixsa\ becomes less sensitive to set-up parameters such as the size of the training sample, the number of inference rounds, even the dimension of the latent space (see Appendix \ref{Appendix C} for an example with real \textit{XMM-Newton} EPIC-PN data). Yet for the  importance sampling correction to be accurate, for a given number of posterior samples (10,000-50,000), 100,000-500,000 log-likelihood estimates are needed, and this may become an important component of the simulation budget. The same code used to generate the spectra generates the \cstat\ and this computation can be easily parallelized within \texttt{PyXspec}, leading to an acceptable run time (about 10 minutes). We have shown, however, that there was a workaround in some cases. Over the range of the approximate posteriors, a neural network can learn the likelihood function, with a limited number of simulations (say 50,000 for \xifu-like spectra). Once the \likelihoodemulator\ has been trained (within at most a couple of minutes), getting hundreds of thousands of approximated likelihoods is done within tens of seconds, enabling the importance sampling correction to be accurate. With the models considered here (and the number of free parameters), we have found that the \likelihoodemulator\ provided sufficient accuracy for the importance sampling correction, whether the weighted importance sampling or the sampling-importance-resampling method, even in the case of complex multi-modal posteriors. The \likelihoodemulator\ will better work when the parameter space is narrow but may face issues when, for instance, the posteriors follow a multi-modal distribution or the number of free model parameters becomes too large. In that case, it is recommended to compute the exact likelihood through extensive simulations, at the expense of dominating the simulation budget. 

Multi-round inference is an iterative process, and its convergence can be precisely tracked through various indicators, such as the shrinkage of the prior proposal towards the observation, the sampling efficiency as defined above and changes in the shape of the posterior distribution. Additionally, as a by-product of reducing the dimensionality of the X-ray spectra, it becomes possible at any round to retrieve the model parameters from the latent representations of the spectra. This can be achieved with a neural network, similar to the \parameterretriever\ presented above. This can help identify the regions of the parameter space that best match the observation. More importantly, this approach informs about the efficiency of the dimension reduction and reveals whether all model parameters can be constrained by the target observation. This information can guide users in determining which parameters can be fixed during the inference process. As a complementary avenue of investigation, leveraging this information to detect model mis-specifications could be explored, although such an analysis is beyond the scope of this paper. 

Following \cite{zhang2023mla..confE..38Z}, it may be worth investigating the sampling efficiency as a stopping criterion for the neural density estimator training. As described above, the surrogate posterior is refined over multiple rounds, with each round using simulations generated from the previous surrogate posterior. These simulations can be used to compute importance weights at no additional computational cost, allowing each round to contribute to the effective sample size. The cumulative sampling efficiency can then serve as an indicator of convergence. Once the sampling efficiency no longer improves, the NDE training can be terminated, and additional effective posterior samples can be obtained via importance sampling from the most efficient surrogate posterior.

Real \xifu\ spectra will not be available until the launch of \textit{NewAthena} in the late 2030s. In the meantime, \xrism-Resolve data provide an additional avenue for validating \sixsa. As shown above, the first test on a Perseus cluster observation has been successful and revealed no limitations, yet provided insights on future improvements. Similarly, real data at lower spectral resolution, such as those from \textit{NICER}, \textit{XMM-Newton}\ldots, enabled probing the technique in various statistical regimes (see Appendix \ref{Appendix B} and \ref{Appendix C}). Across all tests performed to date, \sixsa\ has been shown to deliver posterior distributions identical to those obtained with \bxa.

The ability of \sixsa\ to handle real data, despite being developed on mock \xifu\ data, positions it as a promising tool for the community. Since \sixsa\ relies on the widely used and familiar \xspec\ package for simulations and model access, our next objectives are threefold: (1) to minimize the number of hyperparameters to ensure reliable and reproducible posteriors, (2) to provide diagnostic tools for validating the results independently, and (3) to offer practical guidelines for users. Integration of \sixsa\ within \xspec\ may also be an option.

We are currently handling a handful of hyperparameters from \sbi. As stated above, using a masked auto-regressive flow as the density estimator, with 10 transformations and 100 hidden units each combined with a truncated prior proposal that rejects samples falling outside the $10^{-3}$ quantiles, has shown robust performance across all tested cases. Concerning diagnostics, as discussed in \cite{Dax2023PhRvL.130q1403D}, in addition to providing the (unbiased) Bayesian evidence for comparing models, metrics such as the sample efficiency can be used to assess the proposal quality and identify potential failure cases (see also \cite{Gebhard_2025A&A...693A..42G}). A range of diagnostic tools is also available to check the quality and reliability of posterior inference, five of which are integrated into the \sbi\ package \citep{Tejero2020JOSS....5.2505T,deistler2022pnas2207632119,hermans2022crisis,miller2021truncated,cook2006validation,talts2018validating,linhart2023c2st,lemos2023sampling,schmitt2023detecting}. These tools serve different purposes: for example, model mis-specification checks do not assess the posterior directly but instead evaluate whether the observed data could plausibly have been generated by the model simulated \citep{schmitt2023detecting}. A comprehensive evaluation of the performance and suitability of these diagnostic tools is beyond the scope of this paper.

For initial usage guidelines, we recommend the following setup for \xifu-like or \xrism-Resolve-like spectra: without checking whether the sampling efficiency has reached an acceptable threshold, five rounds of inference with 50,000 simulations per round (doubling this for the first round), a 64-dimensional latent space obtained using a compact auto-encoder with one intermediate layer (containing half as many units as spectral bins), and an additional 100,000–500,000 simulations for importance sampling correction (for typical sampling efficiencies). This conservative setup has proven effective for even the most complex models tested, enabling the generation of exact posteriors in well under one hour per observation when using a \likelihoodemulator\ for the most challenging case. For spectra with lower spectral resolution or less complex models, the number of simulations per round could potentially be reduced by a factor of 10 (5,000), although the time required to generate simulations becomes a minor component of the total runtime. More tests on real data will be required to refine these guidelines. 

Insufficient simulations or poor training of the neural density estimator will become evident during the importance sampling correction phase, leading to low sampling efficiencies. In such cases, the posterior distributions will exhibit broad, diffuse, and irregular shapes with poorly defined peaks (see Appendix \ref{Appendix C}). The contours in the pairwise correlation plots will be spread out and less elliptical. These characteristics are evidence that the inference process has not converged to a well-constrained solution, leaving the parameter space poorly resolved. When this occurs, increasing the training sample size, followed, if necessary, by additional inference rounds is the logical next step. As a sanity check, one should also verify that the validation loss has reached a stable minimum (see Figure \ref{fig:test-case-I-training-history}) and shows no signs of substantial overfitting.

\section{Conclusions}
\label{conclusions}
Applied to X-ray spectral fitting, simulation-based inference that couples auto-encoder-driven dimensionality reduction with likelihood-based importance sampling efficiently yields asymptotically exact posteriors. It also shows that deep-learning inference can meet the accuracy demanded by scientific applications. Notably, it achieves this rapidly and with limited computational resources, even on conventional computers. We consider the ability to limit the energy consumption required for analyzing complex X-ray data a significant advantage of this technique. This is particularly relevant at a time when reducing the environmental footprint of research, including computational resource usage,  has become imperative.


\begin{acknowledgements}
The authors are grateful to the referee for very valuable comments that helped to clarify the content of the paper. They extend their sincere thanks to all colleagues for their invaluable support and encouragement in advancing the potential of SBI-NPE. We are especially grateful to François Mernier for his expert guidance on \xrism-Resolve spectra. In addition to the \texttt{sbi} package \citep{Tejero2020JOSS....5.2505T}, this work made use of many awesome Python packages: \texttt{ChainConsumer} \citep{Hinton2016}, \texttt{keras} \cite{keras_2018ascl.soft06022C}, \texttt{matplotlib} \citep{matplotlib}, \texttt{numpy} \citep{numpy}, \texttt{pandas} \citep{pandas}, \texttt{pytorch} \citep{pytorch}, \texttt{scikit-learn} \citep{scikit-learn}, \texttt{scypi} \citep{scipy}, \texttt{tensorflow} \citep{tensorflow}. To enhance the clarity and readability of the manuscript, the authors, who are not native English speakers, made use of language editing tools, for the sole benefit of the reader.
\end{acknowledgements}

\bibliographystyle{aa} 
\bibliography{dbarret} 

\appendix
\section{Retrieving the model parameters from the latent space}
\label{Appendix A}
\begin{figure}
    \centering
    \includegraphics[width=0.675\linewidth]{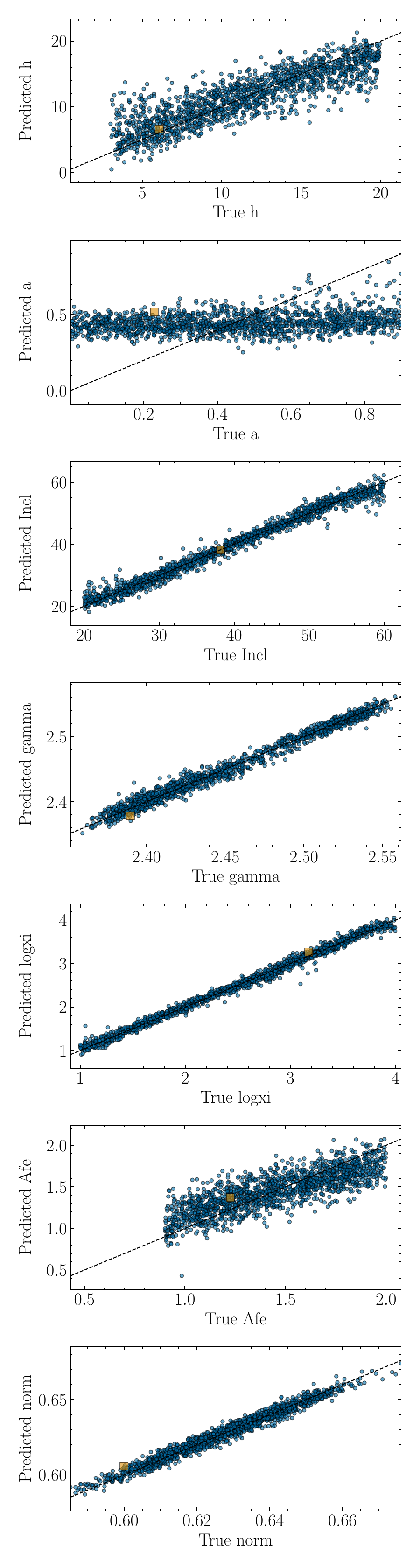}
    \caption{Mapping between the model parameters and the latent space at round 2 to show the sensitivity or lack of the observation to the model parameters. Three parameters can be predicted to have looser constraints. This is particularly the case for the black hole spin and to a lower degree for the irradiating source height and Fe abundance.}
    \label{fig:appI_parameter_mapper_relxillp}
\end{figure}

\begin{figure}
    \centering
    \includegraphics[width=\linewidth]{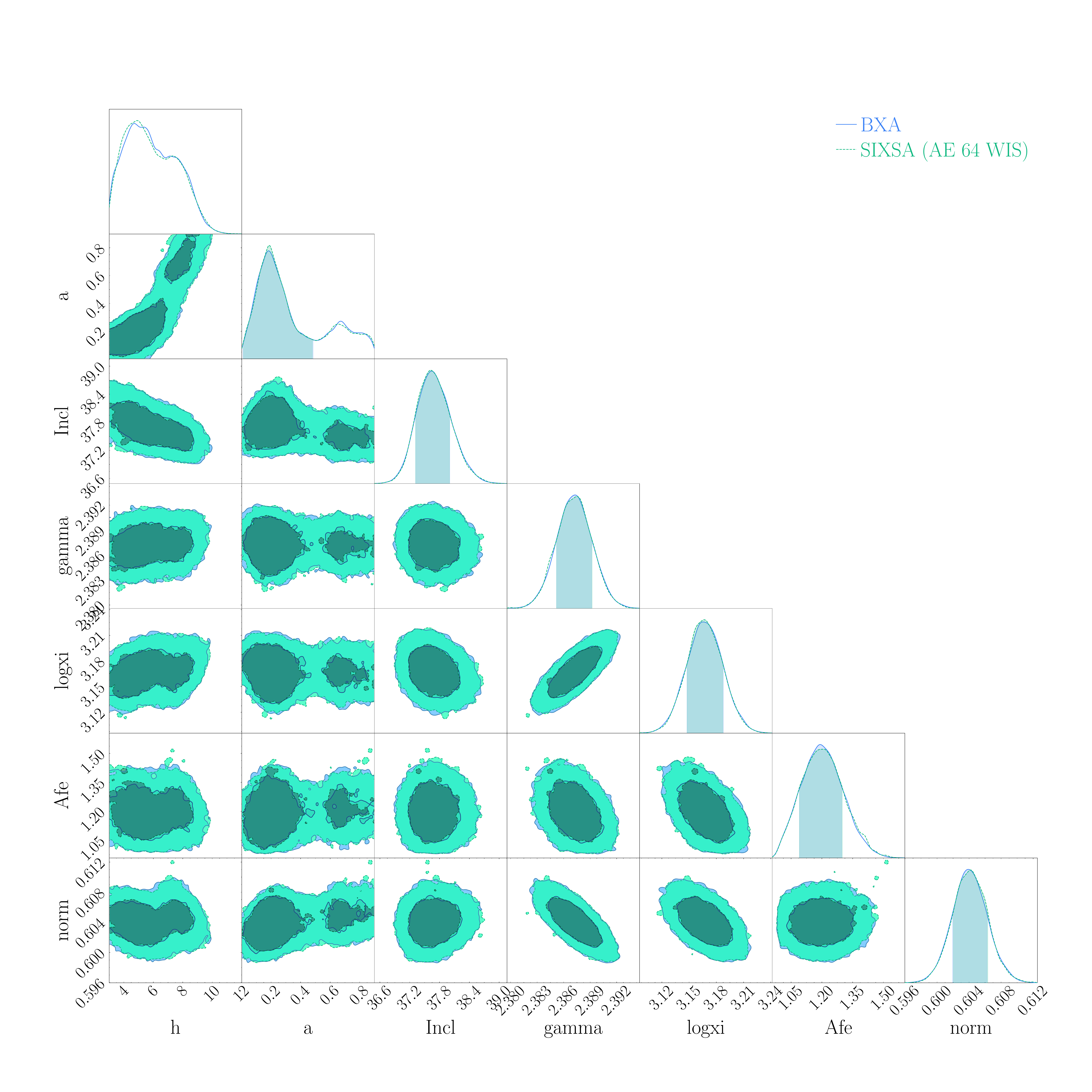}
    \caption{\bxa\ and \sixsa\ posteriors. The \sixsa\ posteriors have been corrected by weighted importance sampling considering 400,000 likelihoods. As expected three out of seven parameters have poorer constraints, most particularly the black hole spin.}
    \label{fig:appI_bxa_sixsa_posterior_relxillp}
\end{figure}

Running the \parameterretriever\ on the latent space provides insights into the sensitivity of a given observation to the model parameters. As an illustration, we consider a low-statistics spectrum from the \texttt{relxillp} model. We perform five inference rounds with 20,000 samples each, except for the first round, where the sample size is doubled. The model includes seven free parameters (see Table \ref{tab:parameter_limits_caseII}). The \parameterretriever\ was applied to the latent space at the second inference round, and its predictions were compared against the true model parameters for the test set (see Figure~\ref{fig:appI_parameter_mapper_relxillp}). A clear correlation is observed for four parameters, while for the remaining three, including the black hole spin, the predictions either partially track or completely fail to reproduce the input values. The corresponding posteriors confirm this behavior (see Figure~\ref{fig:appI_bxa_sixsa_posterior_relxillp}), showing that only four parameters are constrained by the observation, whereas the spin, the irradiating source height, and the iron abundance remain poorly constrained. In essence, if the \parameterretriever\ fails to capture the information related to a model parameter, that parameter cannot be constrained by the observation. While this intermediate step is not required by \sixsa, it provides useful diagnostic insight into the sensitivity of the observation to each free parameter.

\section{Application to \textit{NICER} data}
\label{Appendix B}

 We now turn to \textit{NICER} observations allowing us to probe both high and low statistics regimes and lower resolution X-ray spectra. \textit{NICER} (the Neutron star Interior Composition Explorer, \cite{Gendreau2012SPIE.8443E..13G}) is dedicated to the study of neutron stars through high-throughput, soft X-ray timing and spectroscopy in the 0.2--12~keV range. Equipped with concentrator optics and silicon drift detectors, it achieves an energy resolution of $\sim 85$~eV at 1~keV, comparable for instance to that of \textit{XMM-Newton} EPIC and \textit{Chandra} ACIS, and similar to what will later be provided by the Wide Field Imager onboard \textit{NewAthena} \citep{Antonelli_WFI_2024SPIE13093E..4LA}.

\begin{figure}
    \centering
    \includegraphics[width=\linewidth]{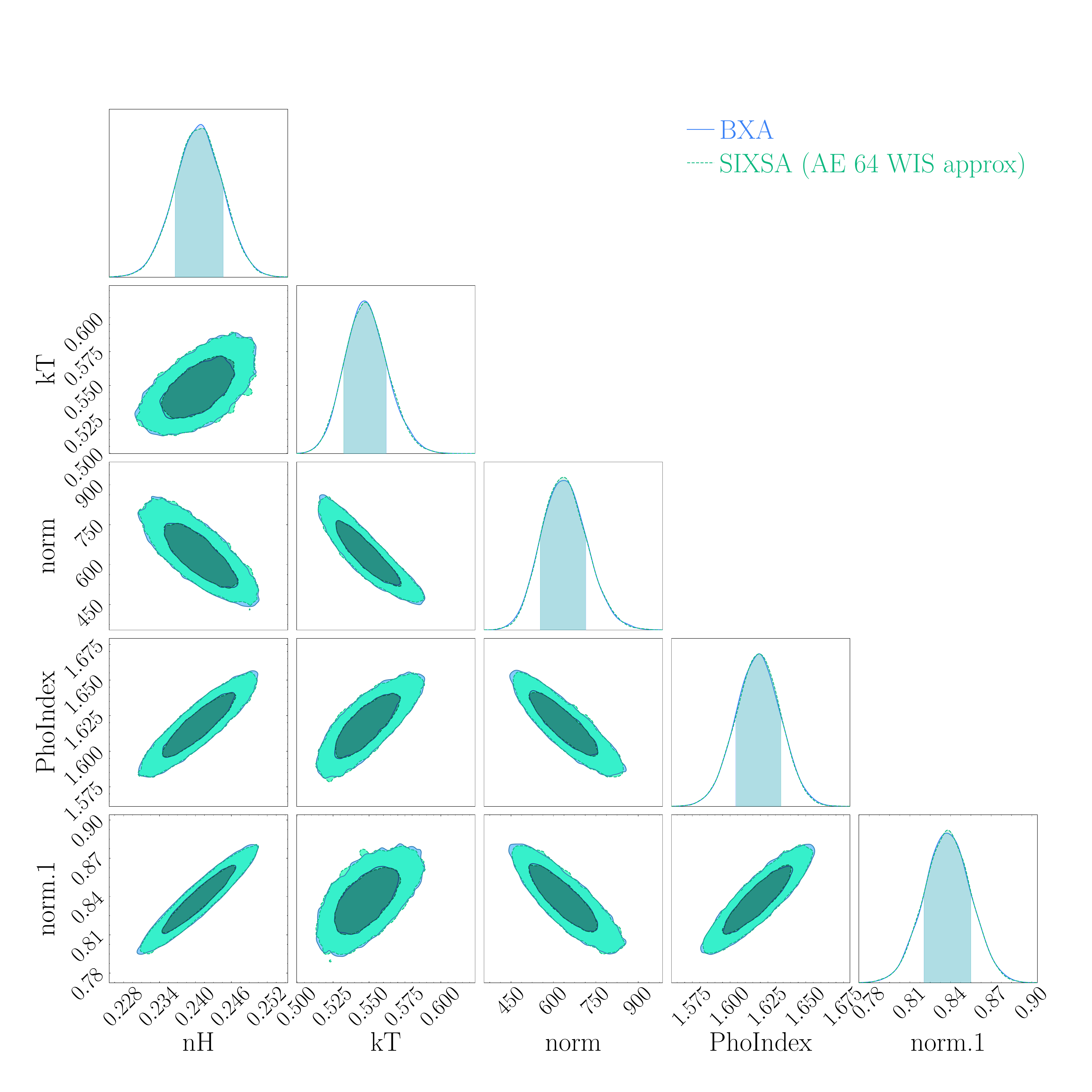}
    \caption{Posterior distributions for both \sixsa\ and the reference provided by \bxa.  The \sixsa\ ones have been obtained after weighted importance sampling, drawing 800,000 samples with the likelihood estimated with the \likelihoodemulator\ trained on 20,000 samples. There is a perfect match between the \sixsa\ and \bxa\ posteriors. The fit applies to a \textit{NICER} spectrum of the persistent emission from 4U1820-303.}
    \label{fig:nicer_persistent}
\end{figure}

In BD24, aiming to demonstrate the working principles of simulation based inference for X-ray spectral fitting, we considered \textit{NICER} data from the X-ray binary 4U1820-303. A \textit{NICER} spectrum has typically 10-20 times less bins than \xifu-like spectra, which is very advantageous in terms of run time, in particular for the simulations (response files are lighter too). We extract a spectrum of the persistent emission of 4U1820-303 for an integration time of 200 seconds. The fit is performed between 0.3 and 10 keV and the spectrum, optimally binned, has more than 200,000 counts over 120 bins. This spectrum enables exploring the Gaussian statistic regime. For such spectrum, modeled as the sum of an absorbed power law plus blackbody, we had a simulation budget of 100,000 spectra for a single round inference and $5 \times 5,000 $ for multiple round inference, and no compression was applied to the spectrum. The training started with a restricted prior eliminating the regions of the initial parameter space the farthest from the observation, based on the \cstat\ (equivalent to adding more inference rounds in our case). This led to posteriors consistent with the ones derived by \xspec\ MCMC. Here we do not consider a restricted prior, instead we extend the prior so that the observation is centered. We keep the training set to 5,000 spectra that we compress into a latent space of 64 dimensions. This ensures that the neural density estimator is well trained, as needed to perform accurate importance sampling. The auto-encoder set up corresponds to a compression factor of $\sim 3$. Although for \textit{NICER}-like spectra, the exact computation of the likelihoods is fast (4,000 simulations per second), we still consider estimating them from the \likelihoodemulator\ that we trained with 10,000 samples. In Figure \ref{fig:nicer_persistent} we compare the \sixsa\ posteriors with the \bxa\ to show the perfect alignment. 

\begin{figure*}
    \centering
    \includegraphics[width=0.485\linewidth]{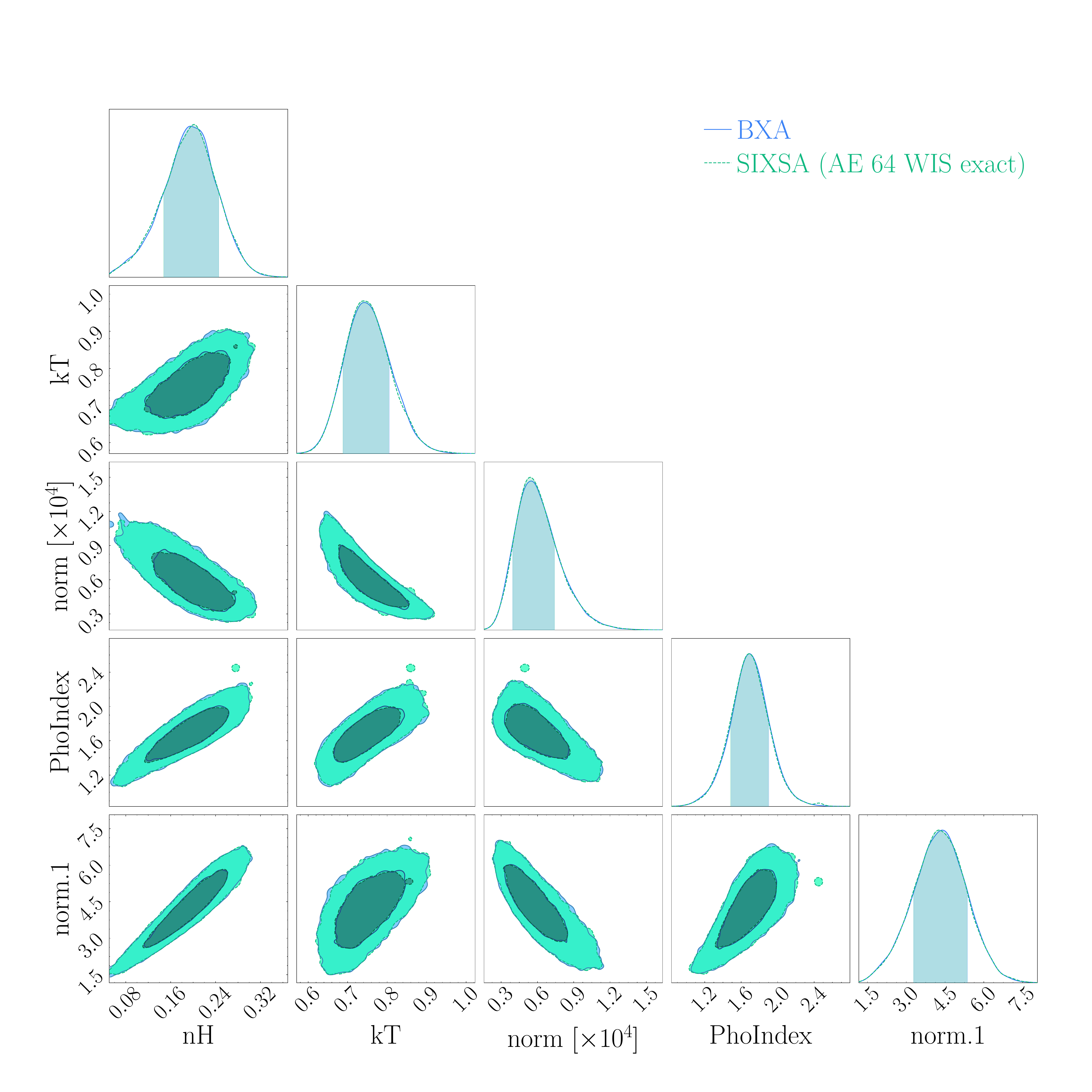}    \includegraphics[width=0.435\linewidth]{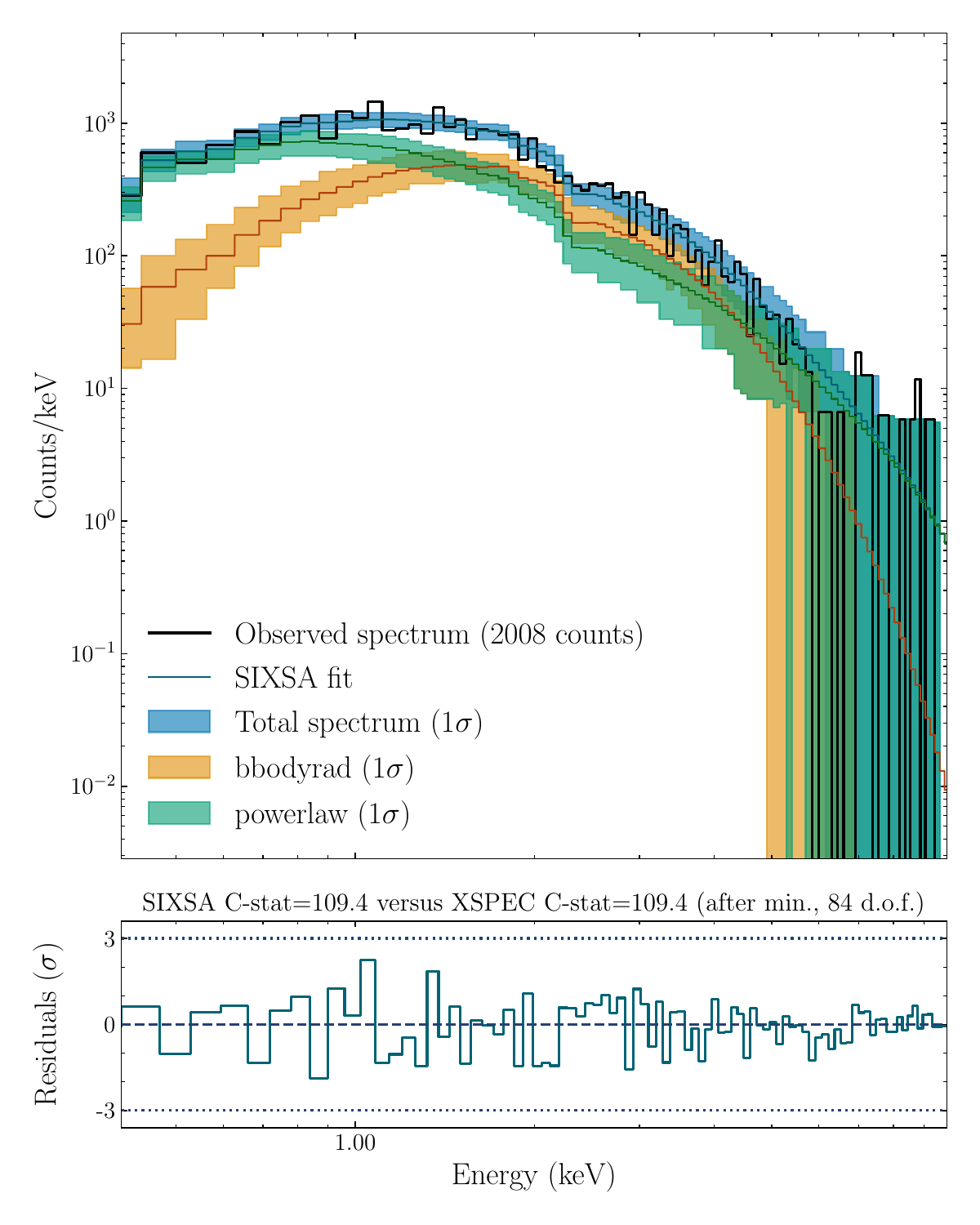}
    \caption{Left: Posterior distributions for both \sixsa\ and the reference provided by \bxa. The \sixsa\ ones have been obtained after weighted importance sampling, drawing 400,000 samples with the exact likelihood estimates. There is a perfect match between the \sixsa\ and \bxa\ posteriors, indicating that the technique works also on real data in the Poisson regime. Right: The folded spectrum and the posterior predictions for the total spectrum and the two components. The spectrum corresponds to a short segment of observation taken at the peak of an X-ray burst: 5 parameters, including the column density, can be constrained with $\sim 2000$ counts.}
    \label{fig:nicer_burst}
\end{figure*}

In BD24 we also explored low count spectra taken by \textit{NICER} along a type I X-ray burst from 4U1820-303. We consider here one of these spectra, optimally grouped, containing $\sim 2,000$ counts over $\sim 90$ bins, exploring the Poisson regime. We fit the spectrum as the sum of an absorbed power law plus a blackbody, leaving the column density free. We run five inference rounds with 5,000 spectra per round, compressed to a 64 latent dimension. We run weighted importance sampling considering 200,000 samples and the exact computation for the likelihoods. We run \bxa\ to get the reference posterior samples. In  Figure \ref{fig:nicer_burst}, we show show the corrected posteriors with the reference \bxa\ ones and the folded spectrum and the 68\% envelope for the model components and their sum. As can be seen the posteriors are identical. We also pushed the latent space dimension to 16 and the results were identical. Given the small number of bins, no compression could be consider. Although there are some signs of overfitting in the latest inference rounds, importance sampling applied to the posteriors derived from training on raw spectra would deliver also exact posteriors. It is interesting to note that projecting the raw spectra to a latent space of dimensions comparable to the number of spectral bins removes the overfitting. This shows the flexibility the method provides in terms of compression factors for the spectra. Starting from the same priors, \sixsa\ derives the exact posteriors in $\sim 4$ minutes, while \bxa\ takes slightly less than one hour. 
\section{Application to \textit{XMM-Newton} data}
\label{Appendix C}
We aim to illustrate the signature of an under-trained neural density estimator. To this end, we analyze the \textit{XMM-Newton} EPIC-PN spectrum of the ultra-luminous X-ray source ULX-4 in NGC 7793, as retrieved by \citet{Quintin2021MNRAS.503.5485Q}. We adopt the same spectral model as in \citet{Dupourque2024A&A...690A.317D}, comprising an absorbed power-law plus a blackbody component. Background is neglected, and we assume a minimum of 20 counts per grouped channel.

\begin{figure}
    \centering
    \includegraphics[width=\linewidth]{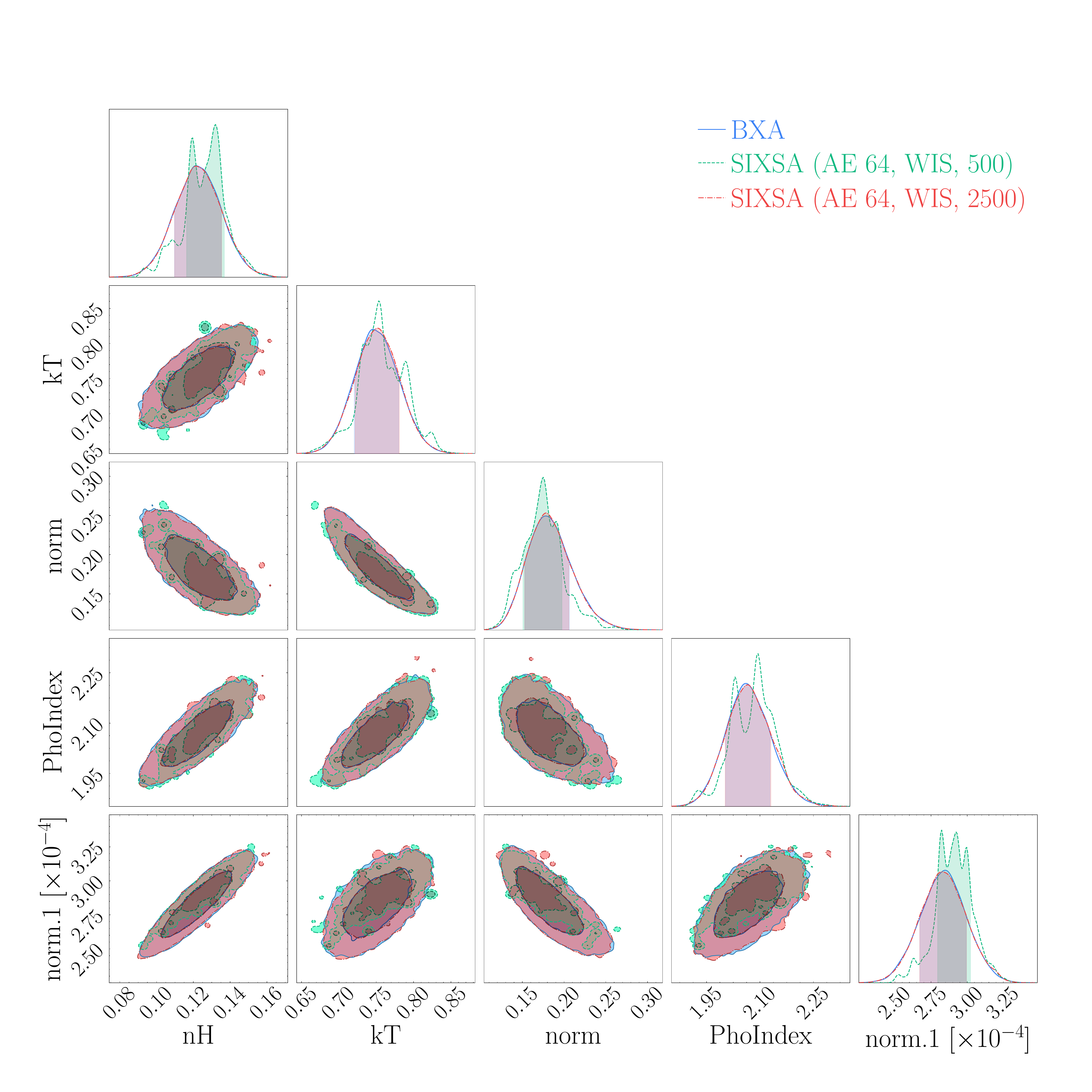}
    \caption{Posterior distributions for both \sixsa\ and the reference provided by \bxa. The \sixsa\ ones have been obtained after weighted importance sampling, drawing 400,000 samples with the exact computation for the likelihood. We consider two training sample sizes for the neural density estimator: 500 and 2,500 samples respectively. The irregular distributions of the 500 run is due to an improper training of the neural density estimator. }
    \label{fig:xmm}
\end{figure}

\begin{figure}
    \centering
    \includegraphics[width=\linewidth]{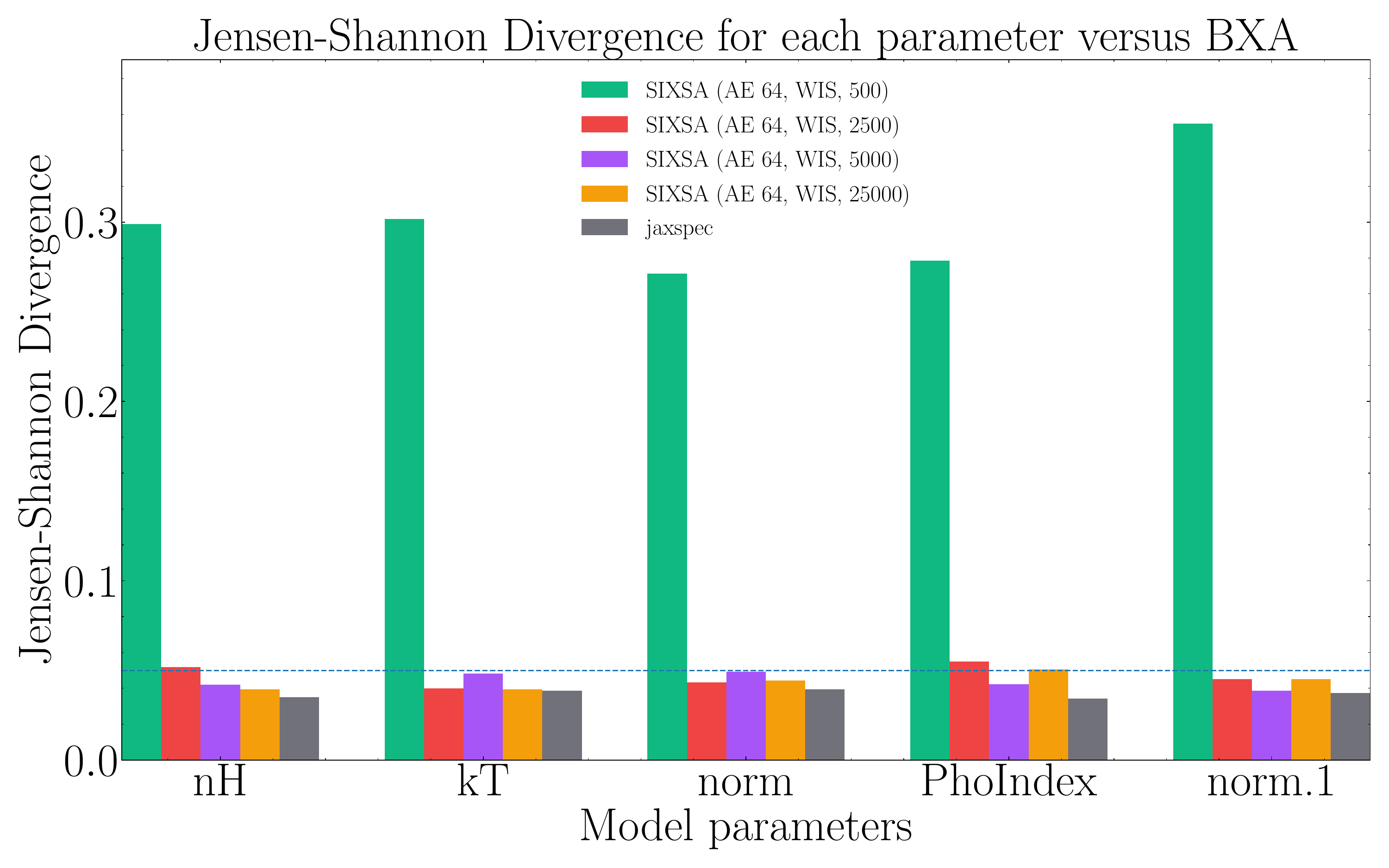}
    \caption{JSD of the \sixsa\ posteriors corrected by weighted importance sampling with respect to the \bxa\ posteriors. Four training sample sizes are considered for the neural density estimator : 500, 2500, 5,000 and 25,000 respectively. The horizontal dashed lines indicate the limit under which the posterior distributions can be considered as identical to the \bxa\ ones \citep{Dax2023PhRvL.130q1403D}. The JSD with the posteriors computed by \jaxspec\ is also shown (a burnin of 1,000 and a chain length of 10,000).}
    \label{fig:xmm_jsd}
\end{figure}

With this setup, we simulate spectra at a rate exceeding 6,000 spectra per second using 16 CPU cores. Our initial training set consists of 500 spectra, roughly an order of magnitude fewer than typically required for reliable inference. We use 5 inference rounds and set the latent space dimensionality of the auto-encoder to 64. Subsequently, we apply weighted importance sampling, evaluating 400,000 likelihoods via exact computation, which is fast due to the high simulation speed.
For comparison, we perform a second \sixsa\ run using a larger training set of 2,500 spectra, still relatively small by recommended guidelines as listed above. Figure \ref{fig:xmm} contrasts the resulting \sixsa\ posteriors with the reference \bxa\ posteriors. Several observations can be made. First, as expected, the use of only 500 training samples leads to poor performance of the neural density estimator, reflected in the irregular, spiky posterior distributions. Second, although 2500 samples remain on the low side, the inclusion of importance sampling significantly improves the results, bringing the posteriors much closer to the \bxa\ reference. This translates into a markedly higher sampling efficiency in the latter case.
To demonstrate the flexibility that importance sampling offers in terms of training sample size for the neural density estimator, we conduct two additional \sixsa\ runs using 5,000 and 25,000 training samples, respectively, while keeping the auto-encoder latent dimension fixed at 64. Figure \ref{fig:xmm_jsd} shows the JSD between the corrected \sixsa\ posteriors and the reference \bxa\ posteriors. As expected, the poorly corrected posteriors from the 500 sample run result in a significantly higher JSD. In contrast, using 2500 samples already yields good performance, with diminishing returns observed beyond that. When analyzing the JSD between the approximate posteriors and the \bxa\ reference, we observe more variability as a function of sample size. However, a general trend emerges: as the number of training samples increases, the approximate posteriors tend to converge toward the \bxa\ results. These findings highlight both the critical role of importance sampling and the flexibility it offers in managing training sample sizes.
\end{document}

%% file: figures/caseI/parameter_limits.tex
\begin{table}[ht]
    \centering
    \begin{tabular}{lcc}
    \hline
    {Parameter Name} & {Free} & Range \\
    \hline
    nH & False & [0.2] \\
PhoIndex & True & $\mathcal{U}(0.5, 4)$ \\
norm & True & $\text{Log-}\mathcal{U}(0.01, 10)$ \\
LineE & False & [2] \\
Sigma & True & $\mathcal{U}(0.005, 0.015)$ \\
Redshift & True & $\mathcal{U}(0.1, 0.5)$ \\
norm & True & $\text{Log-}\mathcal{U}(0.001, 0.3)$ \\
\hline
    \end{tabular}
    \caption{Parameter settings for the model consisting of an absorbed power law and a narrow gaussian line, including the parameter name, the free or frozen status. For frozen parameters, the value of the parameter is given in brackets. For free parameters, the prior type, either uniform $\mathcal{U}$ or log-uniform Log-$\mathcal{U}$, and the range of variation is listed between parenthesis.}
    \label{tab:parameter_limits_caseI}
    \end{table}
    

%% file: figures/caseII/parameter_limits.tex
\begin{table}[ht]
    \centering
    \begin{tabular}{lcc}
    \hline
    {Parameter Name} & {Free} & Range \\
    \hline
    nH & False & [0.02] \\
h & True & $\mathcal{U}(3, 20)$ \\
beta & False & [0] \\
a & True & $\mathcal{U}(0, 0.9)$ \\
Incl & True & $\mathcal{U}(20, 60)$ \\
Rin & False & [-1] \\
Rout & False & [400] \\
z & False & [0] \\
gamma & True & $\mathcal{U}(1.5, 3)$ \\
logxi & True & $\mathcal{U}(1, 4)$ \\
Afe & True & $\mathcal{U}(0.9, 2)$ \\
Ecut & False & [300] \\
refl\_frac & False & [1] \\
switch\_returnrad & False & [1] \\
switch\_reflfrac\_boost & False & [0] \\
norm & True & $\text{Log-}\mathcal{U}(0.01, 10)$ \\
\hline
    \end{tabular}
    \caption{Parameter settings for the \texttt{relxillp} model, including the parameter name and whether the parameter is free or fixed. 
    For fixed parameters, the value is given in brackets. For free parameters, the prior type—either 
    uniform $\mathcal{U}$ or log-uniform $\log\mathcal{U}$—and the range of variation are shown in parenthesis.}
    \label{tab:parameter_limits_caseII}
    \end{table}
    

%% file: figures/caseIII/parameter_limits.tex
\begin{table}[ht]
    \centering
    \begin{tabular}{lcc}
    \hline
    {Parameter Name} & {Free} & Range \\
    \hline
    nH & False & [0.02] \\
kT & True & $\mathcal{U}(0.5, 5)$ \\
He-Ni & False & [1] \\
Redshift & True & $\mathcal{U}(0.05, 0.2)$ \\
Velocity & False & [100] \\
norm & True & $\text{Log-}\mathcal{U}(0.01, 10)$ \\
kT & True & $\mathcal{U}(0.5, 5)$ \\
He-Ni & False & [1] \\
Redshift & False & Tied \\
Velocity & False & [100] \\
norm & True & $\text{Log-}\mathcal{U}(0.01, 10)$ \\
\hline
    \end{tabular}
    \caption{Parameter settings for the double absorbed \texttt{bvapec} model, including the parameter name and whether the parameter is free or fixed. 
    For fixed parameters, the value is given in brackets. For free parameters, the prior type—either 
    uniform $\mathcal{U}$ or log-uniform $\log\mathcal{U}$—and the range of variation are shown in parenthesis. The abundances from He to Ni are set to their default values. }
    \label{tab:parameter_limits_caseIII}
    \end{table}
    